\documentclass[article]{IEEEtran}
\usepackage{graphicx}
\usepackage{grffile}
\usepackage{setspace}
\usepackage{longtable}
\usepackage{url}
\usepackage{cite}
\usepackage{multirow}
\usepackage{tabularx}
\usepackage{amsmath}
\usepackage{subcaption}
\usepackage{helvet}
\usepackage{algorithm,algpseudocode}
\usepackage[usenames, dvipsnames]{color}

\title{TSN Algorithms for Large Scale Networks: A Survey and Conceptual Comparison
\thanks{Please direct correspondence to M.~Reisslein (reisslein@asu.edu).}
}

\author{Ahmed Nasrallah, Venkatraman Balasubramanian, Akhilesh Thyagaturu, Martin Reisslein, and Hesham ElBakoury
\thanks{A.~Nasrallah, V.~Balasubramanian, A.S.~Thyagaturu, and
  M.~Reisslein are with the School of Electrical, Computer, and Energy
  Eng., Arizona State University, Tempe, AZ 85287-5706, USA, Phone:
  480-965-8593, Fax: 480-965-8325, (e-mail: \{ahnasral,
  vbalas11@asu.edu, athyagat, reisslein\}@asu.edu).}
\thanks{H.~ElBakoury is with Futurewei Technologies Inc.,
2330 Central Expressway, Santa Clara, CA, 95050, USA
    (e-mail: helbakou@futurewei.com).} }

\begin{document}

\maketitle

\begin{abstract}
This paper provides a comprehensive survey of queueing and
scheduling mechanisms for supporting large scale deterministic
networks (LDNs). The survey finds that extensive mechanism design
research and standards development for LDNs has been conducted over the past
few years. However, these mechanism design studies have not been
followed up with a comprehensive rigorous evaluation. The main outcome
of this survey is a clear organization of the various research and
standardization efforts towards queueing and scheduling mechanisms for
LDNs as well as the identification of the main strands of mechanism
development and their interdependencies. Based on this survey, it
appears urgent to conduct a comprehensive rigorous simulation study of
the main strands of mechanisms.
\end{abstract}

\section{Introduction}

\subsection{Motivation}
Traditional services for packet switched networks involved a best
effort process that handled and adequately ensured average
latencies. Typically, different methods exist (e.g., explicit
congestion notification, flow and congestion control) that feeds back
notifications to control and to ``slow'' down the data rates of
different applications. This control ensures network stability and
overall fairness and inter-operation between different stream/flows in
a network of many disparate applications operating on the common
converged Ethernet technology networks. However, throttling data rates
due to network fluctuations is not an option for real-time
applications (e.g., cyber-physical systems) due to the nature of these
applications~\cite{finn2018introduction}. Moreover, accurately
determining upper latency bounds, and guaranteeing zero packet loss,
and minimal jitter (delay variation) is severely limited in
traditional networks. Traditional networking technologies can provide
these guarantees typically only with sophisticated highly engineered
middle boxes over small scale networks. A deterministic forwarding
service is highly desirable for strict real-time applications that
enables the convergence of Information and Operational Technology
(IT/OT) under a unified Ethernet technology.

The IEEE 802.1 TSN working group (evolved from Audio/Video Bridging,
AVB group) is working to develop Time Sensitive Networking (TSN)
standardizations that target deterministic forwarding applications and
bridging between layer 2 networks. Specifically, applications that
involve (in addition to multimedia) industrial control, automotive,
and avionics applications, and mobile backhaul that require
just-in-time delivery of data traffic. Similarly, the IETF Deterministic
Networking (DetNet) group is working in collaboration with the TSN
group to develop standardization of IP (L3) layer deterministic
forwarding services.

The main enabler for synchronous traffic deterministic forwarding
services is the Cyclic Queuing and Forwarding (CQF) protocol.  More
specifically, the CQF protocol typically combines a Time-Aware Shaper
(TAS) at the egress port of a switch and Per-Stream Filtering and
Policing (PSFP) at the ingress port of a switch, to shape and regulate
the transmission selection.  This transmission selection within
switches utilizes time division multiplexing based on an underlying
time synchronization in L2 bridged
networks~\cite{messenger2018time}. The CQF protocol results in delays
that are a function of the Cycle Time (CT),
which is typically set according to the Quality of Service (QoS)
characteristics of all Scheduled Traffic (ST) flows, and the number of hops.

In this paper, we provide a comprehensive survey of the various
scheduling (forwarding) mechanisms for ensuring deterministic QoS in
large scale networks.
Following~\cite{qiang-detnet-large-scale-detnet-04}, we define a
large-scale network as a network that covers a large geographic area
so that there are long propagation delays between network nodes and
switches; in particular, a large scale network is a multi-hop network
with long propagation delays between adjacent switches. Moreover, a
large-scale network has a large number of network nodes and switches
as well as a large number of traffic flows. More specifically, a
large-scale deterministic network (LDN) is a large-scale network with
a large number of traffic flows requiring deterministic quality of
service.  Most scheduling mechanisms for LDNs are variations of
underlying CQF scheduling principles. We comprehensively survey these
variations of CQF as well as their advantages, as well as shortcomings
and limitations.  We outline the implication for future research.

\subsection{Contributions}
Our main contribution is in the form of a comparative analysis of
forwarding protocols for LDNs.
\begin{itemize}
\item[i)] We outline the main considerations in designing and applying
  a TSN/DetNet forwarding mechanisms for large scale DetNets (LDNs).
\item[ii)] We survey DetNet related standards and other information
  sources for approaches used to guarantee DetNet and TSN QoS
  requirements for LDNs.
\item[ii)] We present a comparative analysis between the main
  proposals for LDNs and highlight the advantages and limitations for
  each.
\end{itemize}

\subsection{Organization}
Section~\ref{sec:back} provides background on the current state of the
TSN and DetNet developments. Sections~\ref{sec:ldn} and~\ref{res:sec}
present the surveyed proposed protocols from standards and from
academic research.  These efforts typically start from CQF and continue
to derivatives of CQF. Throughout, we discussion and compare the
presented approaches.  Finally, Section~\ref{sec:concl} concludes the
paper.

\section{Background: IEEE 802.1 Time Sensitive Networking (TSN) and Deterministic Networking (DetNet)}
This section provides a brief overview of the standards and research
in TSN and DetNet that are relevant to large scale networks. TSN has
evolved from AVB due to the growing demands industrial applications,
e.g., Internet of Things (IoT) and Industry 4.0. TSN promises to
provide flows or streams, which are sequences of data packets
belonging to an end-to-end communication between a talker (sender) and
listeners (receivers), with Ultra Low Latency (ULL) with bounded
delays, zero congestion packet loss, and very small jitter.  These
so-called scheduled traffic (ST) flows that receive the ULL service
may coexist with best effort traffic flows. AVB started with the
enhanced clock synchronization (IEEE 1588v2),
802.1Qat~\cite{IEEE8021Qat}, and 802.1Qav~\cite{IEEE8021Qav}, stream
reservation protocol (SRP) and credit-based shaper (CBS),
respectively. Due to the growing popularity and success of these
protocols in professional audio/video production, research and
standardization into determinism started to grow in tandem, prompting
the IEEE 802.1 TSN group to start developing real-time Ethernet
standardization for industrial and automotive applications in L2 and
IETF DetNet in L3. Several standards have already been published,
including, $i)$ Frame Preemption (802.1Qbu and
802.3br)~\cite{IEEE8021Qbu,IEEE8023br}, $ii)$ Time-Aware Shaper
(802.1Qbu, TAS)~\cite{IEEE8021Qbv}, $iii)$ Per-Stream Filtering and
Policing (802.1Qci, PSFP)~\cite{IEEE8021Qci}, $iv)$ Cyclic Queuing and
Forwarding (802.1Qch, CQF)~\cite{IEEE8021Qch}, and $v)$ SRP
enhancements and configuration management
(802.1Qcc)~\cite{IEEE8021Qcc}. A more thorough survey of TSN standards
and research along with DetNet has been provided
in~\cite{nas2019ult}. We briefly review the most popular shapers in
use by TSN in the following subsection.

\subsection{TSN Shapers}  \label{TSNShapers:sec}
\subsubsection{Credit-Based Shaper (CBS)}
The IEEE 820.1Qav (CBS)~\cite{IEEE8021Qav} standard was first
introduced in the AVB group that targeted professional audio/video
applications. CBS operates by utilizing \textit{credits} to transmit
traffic from a particular managed queue (e.g., Stream Reservation (SR)
Class A and B). Two main parameters are used to shape and regulate the
CBS traffic, \textit{sendSlope} and \textit{idleSlope}. Frames are
transmitted when the channel is free and the credit is greater than or
equal to zero. If the channel is busy, then the SR class A/B is queued
and credit is increased by \textit{idleSlope}. When the frame is
transmitted, the credit value decreases by \textit{sendSlope}. If any
additional frame is waiting after a CBS frame is transmitted, then it
can be transmitted (back to back) if the credit is greater than or
equal to zero. Otherwise, it is queued while the credit is increased
by \textit{idleSlope} and other traffic can use the channel. This
effectively spreads out the CBS traffic and avoids traffic burstiness
that could cascade downstream to cause latency and jitter
problems. Note that the SRP~\cite{IEEE8021Qat} is used in conjunction
with CBS to register and reserve available bandwidth.

\subsubsection{Time-Aware Shaper (TAS)}
The IEEE 802.1Qbv (TAS)~\cite{IEEE8021Qbv} standard proposes to
emulate time-division multiplexing for all the queues at a switch port
egress using timed gates that open/close according to a prescribed
schedule, allowing frames full access to the egress link with zero
interference from other queues. An open/close instruction is referred
to as a Gate Control Entry (GCE) that dictates which queues to allow
access to the transmission medium.  The entire cyclic sequence of GCEs
is referred to as the Gate Control List (GCL) which is configured by a
network administrator or a central management entity. After a sequence
of a GCL for a particular switch egress port finishes, it repeats
starting from the first GCE again. Each GCE is opened for a limited
time governed by the window time (or slot time) with a desired
transmission selection algorithm once the queue gate is opened.  For
an example system with two traffic classes, there are commonly two
types of GCEs, one for high-priority Scheduled Traffic (ST) and
another one for low-priority Best Effort traffic (BE).  The ratio of
the ST window duration to the BE window duration in a cycle time
governs the prioritization level of the ST traffic.

\subsubsection{Asynchronous Traffic Shaper (ATS)}
The ATS shaper~\cite{IEEE8021Qcr} is based on the Urgency-based
Scheduler (UBS) by Specht et al.~\cite{Specht2016}. The UBS
follows a per-flow shaped queueing scheme under the flow's defined Ethernet PCP
priority and another subsequent set of FIFO shared queues using
Internal Priority Values (IPV) assigned after the first set of per-hop
per-flow shaped queues. Essentially, ATS uses a mixture of both
per-flow and per-class queuing scheme (more on that in
Section~\ref{queue:sec}), where the scheduler assigns eligibility times
according to a token bucket shaper (Token Bucket Emulation, TBE) that
limits the output rate of flows preventing burstiness at downstream
nodes. ATS operates without time synchronization among nodes in the
network. A more detailed comparison between ATS and TAS is given
in~\cite{nas2019per} and an end-to-end latency analysis of ATS has been
presented in~\cite{bou2018the,mohammadpour2018end,moh2019imp}.

\subsection{Deterministic Networking (DetNet)}
The DetNet architecture~\cite{ietf-detnet-architecture-13} provides
deterministic and reliable forwarding services in Layer 3 (L3). The
DetNet services depend on $i)$ discovering, configuring, and
allocating network resources for DetNet (or TSN) flows, $ii)$
coordinating and orchestrating service and transport functions so that
all DetNet and non-DetNet flows can have a fair share of the
transmission medium, and $iii)$ controlled behaviors of the allocated
resources (including transmission selection process) so that latency
bounds can be guaranteed (similar to the Integrated Services,
IntServ~\cite{rfc2210}, networking model) without over-provisioning
the network resources.

\subsection{Queuing Model in TSN and DetNet}	\label{queue:sec}
In a TSN or DetNet queue management scheme, two broad categories that
handle queuing delay bounds that shape or regulate traffic belonging
to highly prioritized class of service have been
discussed~\cite{finn-detnet-bounded-latency-03}, namely $i)$ per-class
queuing, and $ii)$ per-flow queuing. Per-class queuing is the
traditional method of assigning priority (e.g., in form of the
Priority Code Point, PCP) that corresponds to the class of service
(e.g., Differentiated Service, DiffServ~\cite{rfc7657} in L3). Note
that in Ethernet, $8$ queues per port are used to handle classes of
service depending on the $3$-bit of the PCP value in the 802.1Q
tag. This provides a course-grained QoS to applications that require
low latencies. However, since all the flows share the same PCP value
or class of service, they are queued in the same queue which can
increase the burstiness that then cascades downstream to other flows
sharing the same queue.

An alternative to per-class queuing is per-flow queuing which provides
a queue to every flow traversing a switch port. This per-flow queueing
allows switches to guarantee QoS by managing the flow burstiness and
rate, i.e., Traffic Specification (T-spec) which is used in Integrated
Service (IntServ)~\cite{rfc2210} delivering fine-grained QoS. However,
per-flow queuing increases complexity and cost as the network scales
up which is not practical in many scenarios. A hybrid approach
involves using per-class queuing with interleaved regulators situated
before the queuing subsystem within a switch similar to the 802.1Qcr
(ATS)~\cite{IEEE8021Qcr}. Note that this regulator does not increase
the worst-case delay of the queuing subsystem.

\subsection{Time Synchronization Considerations for Wide Area Networks (WANs)}
The Network Time Protocol (NTP)~\cite{mills1991internet} has been very
successful in ensuring time synchronization in WAN/LANs within a
$10~ms$ margin in most cases. However, as pointed out by Huston et
al.~\cite{Huston2018time}, almost half of the Internet connected
devices are either running fast or slow when compared to the
Coordinated Universal Time (UTC) reference time within a $2$ second
time margin. This problem can be attributed to two main factors: $i)$
clock skew/drift by local machines, and $ii)$ configuration problems
on local clock dates and time. Therefore, WAN-scale networks (which
are typical for DetNet) have to take potential shortcomings in time
and frequency synchronization into account. Ideally, synchronous
traffic shapers should have their transmission scheduling aligned
correctly to ensure that TSN and DetNet flows QoS are maintained.

\section{Standards Efforts: Traffic Shapers, Deterministic Networking, and Cyclic Queuing and Forwarding for Long Distance and Low-Latency Communications} \label{sec:back}

\subsection{Long Range Queuing and Forwarding}	\label{sec:ldn}
\begin{figure}[t!]  \centering
\includegraphics[width=3.5in]{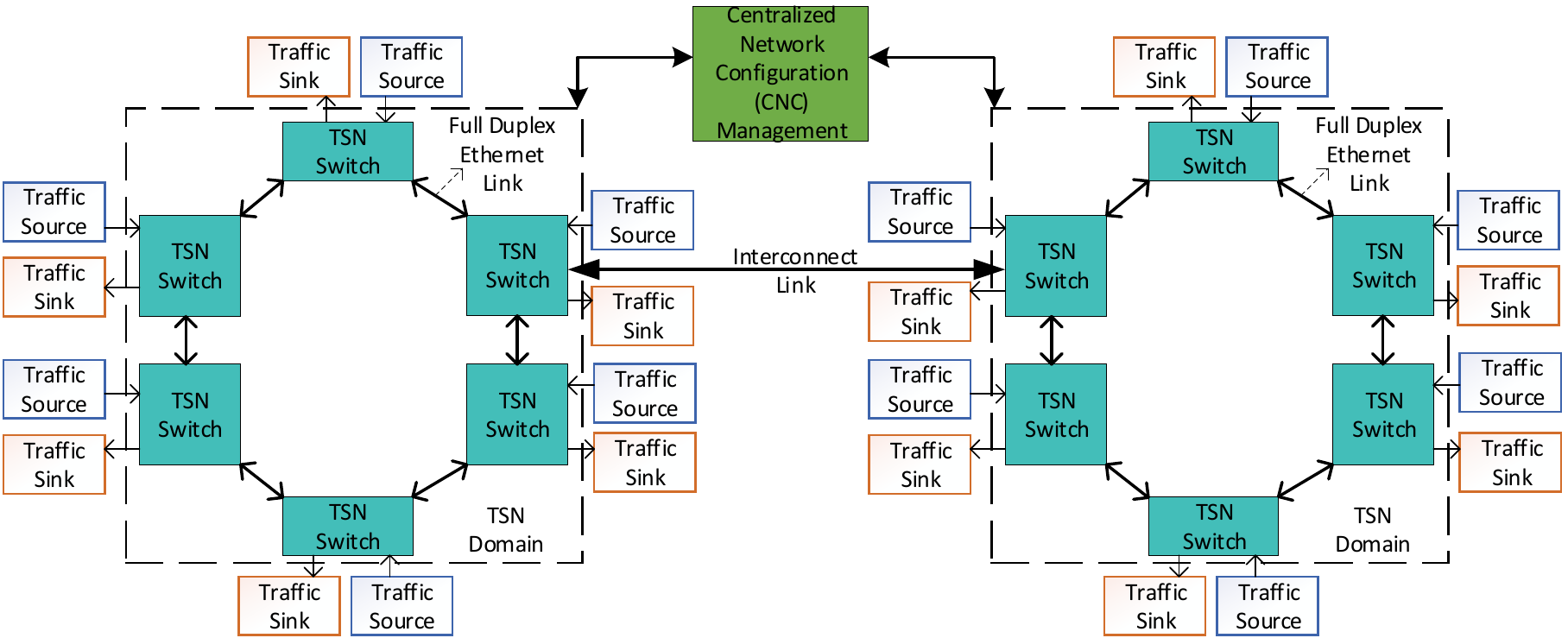} \vspace{-0.5cm}
\caption{Illustration of the rings topology using a high bandwidth
  interconnect link between adjacent rings. Each ring is controlled by
  a single 802.1Qcc Central Management Entity (CNC) with out-of band
  signaling to each switch under the TSN domain highlighted by the
  dashed box around the rings.}
	\label{fig_tsn_longRange}
\end{figure}

\begin{figure}[t!]  \centering
\includegraphics[width=3.5in]{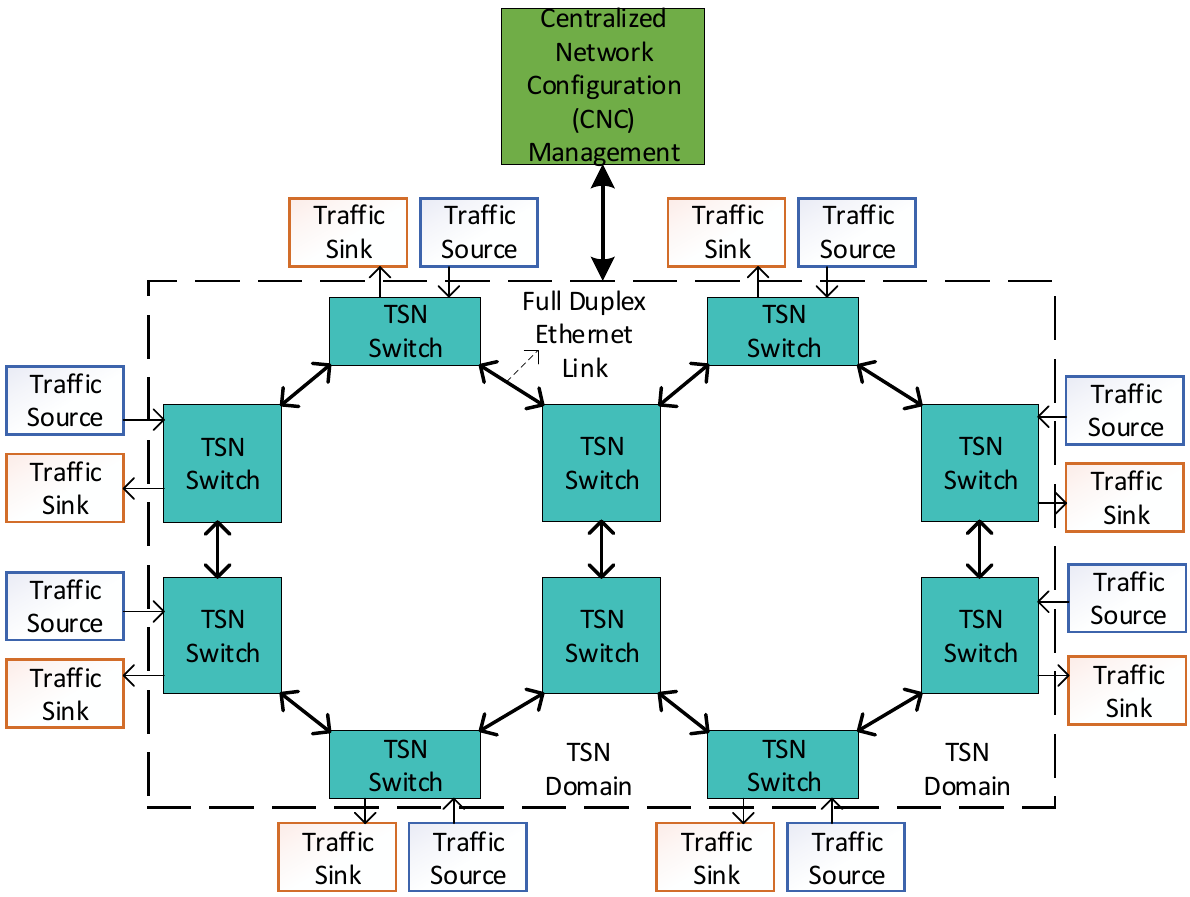} \vspace{-0.5cm}
\caption{Illustration of the rings topology using dual switches to
  connect the rings.}  \label{fig_tsn_longRange_v2}
\end{figure}

\begin{figure}[t!]  \centering
\includegraphics[width=3.5in]{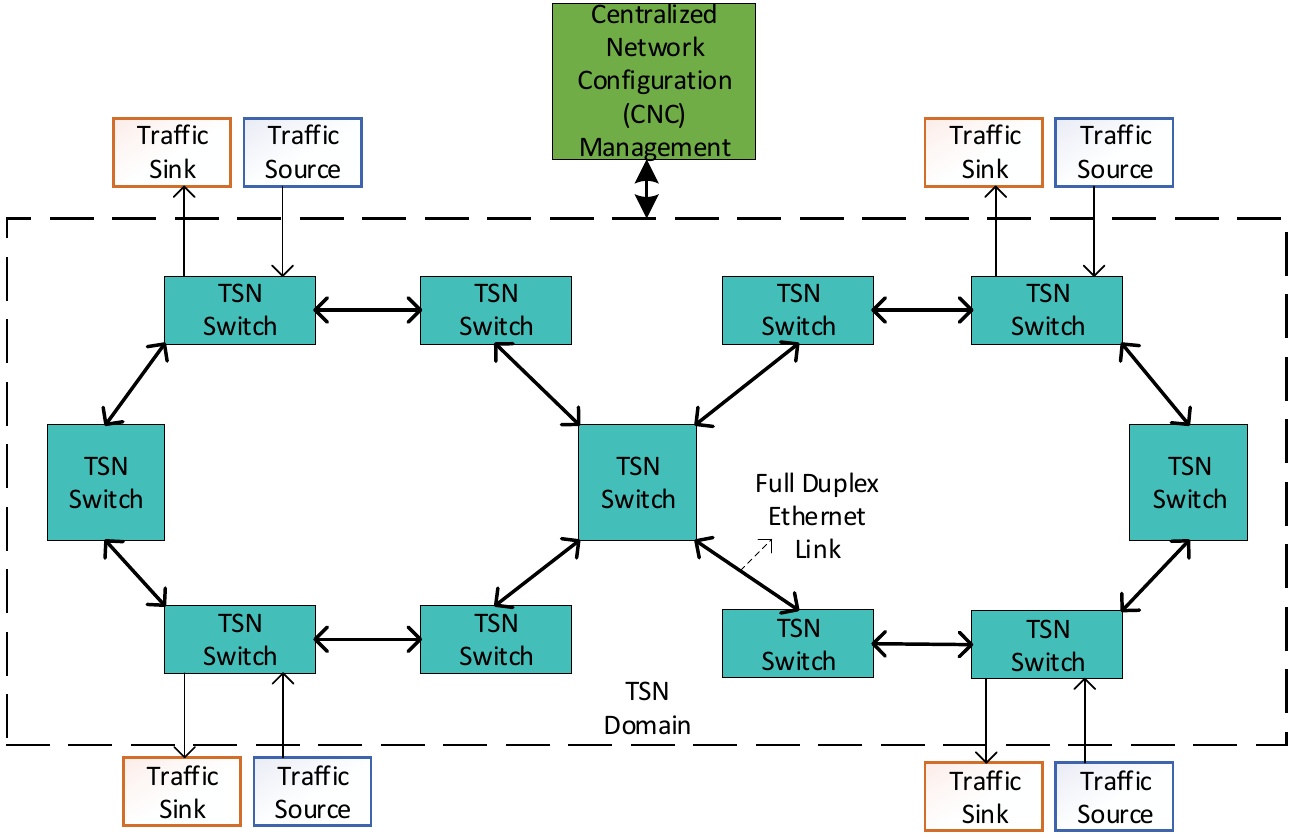} \vspace{-0.5cm}
\caption{Illustration of the rings topology using single switch to
  connect the rings.}  \label{fig_tsn_longRange_v3}
\end{figure}

Generally, TSN is deployed on relatively small scale LANs where the
number of flows is small, the distance between devices is short, and
the number of devices is less than deployments of large scale DetNet
(LDNs) in IP. Thus, the following main challenges need to be considered
when employing TSN principles for large-scale
networks~\cite{qiang-detnet-large-scale-detnet-04}
\begin{enumerate}
\item Limited time synchronization among nodes
\item Generally long propagation delays that can introduce large jitter
\item Per-flow queuing schemes are not scalable due to large state space.
\end{enumerate}
Such LDNs could be modelled with interconnected rings as illustrated in
Figs.~\ref{fig_tsn_longRange}--~\ref{fig_tsn_longRange_v3}.

Assuming that time or frequency synchronization is possible in LDNs,
the main challenge to guarantee TSN QoS for all registered flows
(using any established reservation protocol) is due to the long
propagation delays between adjacent switches along a multi-hop path
from talker to lister. Note that each switch/node in the network can
be configured to operate with per-class or per-flow queuing. Several
draft and published standards have been proposed which are discussed
in the following subsections.

\subsubsection{Cyclic Queuing and Forwarding (CQF)}
The published IEEE 802.1Qch (CQF)~\cite{IEEE8021Qch} standard proposes
to coordinate enqueue/dequeue operations within a switch in a cyclic
fashion.  Groundwork on CQF, which was also previously known as
Peristaltic shaper, was conducted by Thangamuthu et
al.~\cite{Thangamuthu2015}.  Moreover, Thiele et
al.~\cite{thiele2015timeaware} have conducted a theoretical analysis
of the blocking factors for CQF and TAS.

The CQF cyclic operation results in an easily calculable latency
bound governed by the chosen Cycle Time and the number of end-to-end
hops between communicating parties. In CQF, time is divided into slots
or intervals (similar to TAS). For a given traffic class, two queues
are used to enable the cyclic property. Frames arriving in interval
$x$ will be transmitted in interval $x+1$. Similarly, frames arriving
in interval $x+1$ are transmitted in interval $x+2$, and so on.  The
maximum and minimum frame delay bounds in CQF with $H$ and $CT$
representing the number of hops and cycle time duration, respectively,
are
\begin{eqnarray}	\label{eq1}
D_{Max} &=& (H+1) \times CT  \\
D_{Min} &=& (H-1) \times CT  	\label{eq2}
\end{eqnarray}
Two queues are used to handle enqueue and dequeue operations in
separate time intervals. For example, frames arriving in even
intervals will be enqueued in one queue, while the frames that were
enqueued during the previous interval will be transmitted from the
other queue. In CQF, a frame sent by an upstream switch in cycle $x$
must be received by the
downstream at cycle $x$, i.e., the propagation delay must be less than
the selected cycle time. Therefore, the cycle time is constrained by
the link distance (network scale in general). Essentially, the smaller
the network size, the easier it is to guarantee the TSN QoS by
CQF. Additionally, CQF has a few challenges that limit its viability,
such as $i)$ accurately determining the appropriate cycle time, and
$ii)$ cycle duration misalignment where due to processing and
transmission delays, a frame can be received in the wrong cycle (i.e.,
be placed in the wrong outbound queue).

\subsubsection{CQF 3-Queue}
The standard CQF~\cite{IEEE8021Qch} has limited scalability for large
networks and suffers if frames arrive in the wrong cycle. This
prompted the IETF to formulate a draft discussing these issues along
with potential solutions for LDNs. The IETF draft by Finn et
al.~\cite{finn-detnet-bounded-latency-03} presents an analysis and
parameterized timing model on bounded latency for DetNet.  The IETF
draft in-cooperating TSN mechanisms and discusses specifically the
CQF protocol to guarantee bounded delays in DetNet using either TAS or
ATS traffic shapers. Generally, the evaluation of end-to-end latency bounds on
a single DetNet transit or relay node involves several elements within
the node (transmission, propagation, preemption, processing,
regulation, and queuing). Aside from the regulation and queuing
delays, the other delays are highly depended on the hardware and
technology of the node and not on the traffic specification that is
negotiated by the reservation and registration procedure for the
DetNet or TSN flow. When the total per-hop delay of one DetNet packet
transmitted from an upstream node to a downstream node exceeds the
allotted cycle time (e.g., if upstream node sends the packet closer to
the next cycle, and the downstream processing delay is not necessarily
known to the upstream node), a third packet queue (buffer) is needed.

\subsubsection{Scalable Deterministic Forwarding (SDF)}
SDF is currently an IETF
draft~\cite{qiang-detnet-large-scale-detnet-04} within the DetNet
group that proposes to add cycle identifiers to packets traversing
LDNs that operate in a similar manner as CQF. Essentially, each node
(router/switch) has synchronized frequency (not time) and forwards
traffic in a slotted manner according to the cycle identifier carried
in the packet header. Note that the forwarding mechanism can be
asynchronous between neighbor nodes. The cycle identifier is used to
eliminate the time synchronization requirement among nodes. Each
ingress and egress gateway port has a gate function (similar to PSFP)
that shapes or regulates traffic flows, i.e., implements per-flow
queues only at the ingress or egress gateways closer to the
talker/listener instead of the core network. SDF maintains $3$ queues
for a given traffic class at a given port. One queue that dequeues, and
$2$ queues that enqueue (from different cycle identifiers) received packets
designated towards the outbound port. Each packet header carries the
cycle identifier. Therefore, even if two packets are received from an
upstream node during a single cycle but from two different cycles, the
explicit cycle identifier can instruct the downstream node on which
cycle to forward the packet. Each node also maintains a cycle mapping
relationship table that maps incoming packets with a cycle identifier
to the another cycle designated towards the outbound port. These
tables can be configured using centralized (e.g., SDN~\cite{ami2018sdn,biz2016sdn,kel2019ada}) or distributed
control or orchestration
method. The upper latency bound for SDF is similar to CQF
with a per-hop delay of $2 \times CT$ or end-to-end latency of $2
\times CT \times H$, where $H$ denotes the total hop count.

\subsubsection{Cycle Specified Queuing and Forwarding (CSQF)}
Cycle Specified Queuing and Forwarding (CSQF) is proposed in an IETF
draft by Chen et al.~\cite{chen-detnet-sr-based-bounded-latency-01}
within the DetNet group. CSQF leverages Segment Routing~\cite{rfc8402}
Identifier (SID) to coordinate cyclic transmission times across the
LDN offering bounded delay and lossless packet service delivery. CSQF
operates similar to CQF, whereby per-class queuing
(Diffserv~\cite{rfc7657} in IP) is utilized, hence network can scale
up easily. Chen et al. argues that the regular CQF condition, that all
packets sent in a single cycle have to arrive and be queued in the
downstream node in the same cycle, limits the bandwidth utilization
since some bandwidth has to be reserved as a guard band at each
cycle. CSQF improves CQF by explicitly specifying the transmission
cycles at each DetNet node across the entire path from sender to
receiver. For each traffic class of DetNet flows, $3$ queues are
maintained for each outbound port (Sending, Receiving, and Tolerating
queues, denoted as SQ, RQ, and TQ). These roles are not fixed, i.e.,
the queues at each cycle rotate.

\begin{table*}[t!]
	\centering
	\caption{Summary comparison of surveyed standards}
	\begin{tabular}{l|c|c|c}
	\textbf{Approach}                               & \textbf{Synchronization} & \textbf{Topology} & \textbf{Queuing Scheme}         \\ \hline \hline
		CQF                                     & Time/Frequency  & Small    & Per-class              \\
		CQF 3 Queue                             & Time/Frequency  & Large    & Per-Class or Per-Flow  \\
		Scalable Deterministic Forwarding       & Frequency       & Large    & Per-Class and Per-Flow \\
		Cycle Specified Queueing and Forwarding & Frequency       & Large    & Per-Class              \\ \hline
	\end{tabular}
\end{table*}

\subsection{Discussion on Standards Efforts of the Shapers Proposed for LDNs}
A common paradigm that has come to the forefront over the past few
years is the edge data center. More precisely, computing at the edge
reduces latency and works in a more hierarchical manner. From our
initial surveyed LDN approaches, different methods build upon CQF and
attempt to use the cyclic property of CQF as a means of guaranteeing
TSN and DetNet QoS. Usage of CQF in such an environment is primal to
see the benefits of the schedule. Likewise, micro-data centers have
the same usage criteria in terms of time sensitive.  In terms of
research efforts our main criteria to judge each approach includes
\begin{enumerate}
\item Whether the approach uses time, frequency, or no synchronization.
  Essentially, the cost and problems associated with schedule
  preparation need to be evaluated.
\item Amidst all these efforts we envision support for large or small
  scale topologies while utilizing per-class queueing, per-flow queueing, or a mixture of both
  queuing schemes. In the future it appears highly important to
  quantitatively examine
  the impact of CQF in newly emerging networking paradigms,
  such as edge computing and in-network packet cloud computing.
\end{enumerate}

Overall, CQF is emerging to be the main forwarding protocol envisioned
for both TSN and DetNet applications since it provides a simple
analysis of QoS and can be readily integrated using inter-operable
switches so long as the timing between switching reception and
transmission queues is aligned. This assumption is easily violated in
large scale networks that include links with long propagation delays. Also, time
synchronization may generally be limited in large scale networks.

The 3-Queue CQF has been introduced to handle out of synch packets
(due to processing delay variations) that arrive in the wrong cycle by
adding another queue that buffers this type of packets. Generally, if
the switches in a network need to absorb more jitter/burst capacity,
more queues can be used. Similarly, if the network typically has low
jitter and traffic burstiness, then two queues are sufficient.
An open question is whether more than 3 queues are needed in
CQF for LDNs with very long propagation delays.

The IETF proposals (SDF and CSQF) are both very similar in that
they require the packet headers to be augmented with specific
information on which cycle a packet must be transmitted in. Both
operate on the 3-Queue CQF scheme whereby the third queue is used to
absorb jitter and burstiness affected traffic.

A comprehensive rigorous comparison of the various standards
approaches, including 802.1Qch CQF with two and three queues, as well
as SDF and CSQF is missing in the literature.  Such a comprehensive
study should in particular examine the tradeoffs between the time
synchronized approaches (CQF with two and three queues) in comparison
to the other approaches, which can operate without time
synchronization, i.e., SDF and CSQF, across a wide set of operating
conditions.

A related open question is how the regular TSN standard shapers that
have been developed for small-scale networks, namely TAS and ATS
(Section~\ref{TSNShapers:sec}) could be adapted to LDNs.  One strategy
could be to scale up the time bases in these existing standards, e.g.,
to scale up the cycle time in TAS to keep up with the scale up in the
switch-to-switch propagation delays in LDNs, possibly in conjunction
with making TAS more flexible through refinements, see
Section~\ref{ABSASW:sec}.  Future research needs to examine whether
this scaling up of the time bases is a feasible and reasonably
efficient strategy for adapting TAS and ATS to LDNs.  Likely there
are various trade-offs, e.g., increasing the cycle time proportionally
to the link distance will increase the overall delay levels and will
likely waste some transmission resources when switches have no data to
transmit during the extended cycle times. Thus, proper bandwidth reservations
need to be employed throughout.

\begin{table*}[t!] \centering
\caption{Summary comparison of surveyed main approaches.  $\uparrow$
  indicates better, while $\downarrow$ indicates relatively worse. All comparisons assume proper bandwidth reservations.}
	\begin{tabular}{lllllllll}
		\vspace{-0.2cm}	\textbf{Article} & \textbf{Lat.} & \textbf{Overh.} & \multicolumn{1}{c}{\textbf{Compl.}} & \textbf{Flexi.} & \textbf{Cost} & \multicolumn{1}{c}{\textbf{Tput.}} & \multicolumn{1}{c}{\textbf{Depend.}} & \textbf{E.g., App.\vspace{0.3cm}} \\  \hline \hline 
		\multicolumn{9}{c}{\textbf{Traffic Shapers}} \\
TAS ABS~\cite{nas2019per} & \multicolumn{1}{c}{$\downarrow$} & \multicolumn{1}{c}{$\downarrow$} & \multicolumn{1}{c}{$\downarrow$} & \multicolumn{1}{c}{$\uparrow$} & \multicolumn{1}{c}{$\downarrow$} & \multicolumn{1}{c}{$\downarrow$} & \multicolumn{1}{c}{Time triggered} & \multicolumn{1}{c}{General LDNs} \vspace{0.1cm} \\ 

TAS ASW~\cite{nas2019per} & \multicolumn{1}{c}{$\downarrow$} & \multicolumn{1}{c}{$\uparrow$} & \multicolumn{1}{c}{$\uparrow$} & \multicolumn{1}{c}{$\uparrow$} & \multicolumn{1}{c}{$\uparrow$} & \multicolumn{1}{c}{$\uparrow$} & \multicolumn{1}{c}{Time triggered} & \multicolumn{1}{c}{General LDNs} \vspace{0.1cm} \\ 

Seaman~\cite{seaman2019pat}, Paternoster& \multicolumn{1}{c}{$\downarrow$} & \multicolumn{1}{c}{$\uparrow$} & \multicolumn{1}{c}{$\downarrow$} & \multicolumn{1}{c}{$\uparrow$} & \multicolumn{1}{c}{$\uparrow$} & \multicolumn{1}{c}{$\uparrow$} & \multicolumn{1}{c}{Epoch param.} & \multicolumn{1}{c}{General LDNs} \vspace{0.1cm} \\ 

        Ball et. al~\cite{ball2019evaluation} & \multicolumn{1}{c}{$\downarrow$} & \multicolumn{1}{c}{$\downarrow$} & \multicolumn{1}{c}{$\uparrow$} & \multicolumn{1}{c}{$\downarrow$} & \multicolumn{1}{c}{$\uparrow$} & \multicolumn{1}{c}{$\downarrow$} & \multicolumn{1}{c}{Traffic knowl.} & \multicolumn{1}{c}{Smart Grids} \vspace{0.1cm} \\ 
		Specht et al.~\cite{Specht2016} 	& \multicolumn{1}{c}{$\uparrow$} & \multicolumn{1}{c}{$\uparrow$} & \multicolumn{1}{c}{$\downarrow$} & \multicolumn{1}{c}{$\uparrow$} & \multicolumn{1}{c}{$\uparrow$} & \multicolumn{1}{c}{$\downarrow$} &\multicolumn{1}{c}{Timing eval.} & \multicolumn{1}{c}{Audio, Video} \vspace{0.1cm} \\ 
		Mangin et al.~\cite{mangin2019multiplexing}	& \multicolumn{1}{c}{$\downarrow$} & \multicolumn{1}{c}{$\uparrow$} & \multicolumn{1}{c}{$\uparrow$} & \multicolumn{1}{c}{$\uparrow$} & \multicolumn{1}{c}{$\downarrow$} & \multicolumn{1}{c}{$\uparrow$} & \multicolumn{1}{c}{Timing} & \multicolumn{1}{c}{Audio, Video} \vspace{0.1cm} \\ 
		Li et al.~\cite{li2019enhanced}	& \multicolumn{1}{c}{$\uparrow$} & \multicolumn{1}{c}{$\uparrow$} & \multicolumn{1}{c}{$\downarrow$} & \multicolumn{1}{c}{$\uparrow$} & \multicolumn{1}{c}{$\uparrow$} & \multicolumn{1}{c}{$\uparrow$} & \multicolumn{1}{c}{Time Triggered} & \multicolumn{1}{c}{IoT} \vspace{0.1cm} \\ 
		Buratti~\cite{buratti2018joint}	& \multicolumn{1}{c}{$\downarrow$} & \multicolumn{1}{c}{$\uparrow$} & \multicolumn{1}{c}{$\uparrow$} & \multicolumn{1}{c}{$\uparrow$} & \multicolumn{1}{c}{$\downarrow$} & \multicolumn{1}{c}{$\downarrow$} & \multicolumn{1}{c}{Topology Info.} & \multicolumn{1}{c}{Audio, Video} \vspace{0.1cm} \\ 
		Said et al.~\cite{said2019sdn} 	& \multicolumn{1}{c}{$\uparrow$} & \multicolumn{1}{c}{$\uparrow$} & \multicolumn{1}{c}{$\uparrow$} & \multicolumn{1}{c}{$\uparrow$} & \multicolumn{1}{c}{$\uparrow$} & \multicolumn{1}{c}{$\uparrow$} & \multicolumn{1}{c}{SDN} & \multicolumn{1}{c}{Sensors} \vspace{0.1cm} \\ 
		Wetterwald~\cite{wetterwald2018source}	& \multicolumn{1}{c}{$\downarrow$} & \multicolumn{1}{c}{$\downarrow$} & \multicolumn{1}{c}{$\uparrow$} & \multicolumn{1}{c}{$\downarrow$} & \multicolumn{1}{c}{$\uparrow$} & \multicolumn{1}{c}{$\downarrow$} & \multicolumn{1}{c}{Source Info.} & \multicolumn{1}{c}{Audio, Video}  \vspace{0.1cm} \\   \hline  
		\multicolumn{8}{c}{\textbf{Queuing and Forwarding}} \\
		Joung~\cite{joung2019regulating}	& \multicolumn{1}{c}{$\downarrow$} & \multicolumn{1}{c}{$\uparrow$} & \multicolumn{1}{c}{$\uparrow$} & \multicolumn{1}{c}{$\downarrow$} & \multicolumn{1}{c}{$\uparrow$} & \multicolumn{1}{c}{$\downarrow$} & \multicolumn{1}{c}{Classes based} & \multicolumn{1}{c}{Audio, Video} \vspace{0.1cm} \\ 
		Ayub et al.~\cite{ayub2018priority}	& \multicolumn{1}{c}{$\uparrow$} & \multicolumn{1}{c}{$\downarrow$} & \multicolumn{1}{c}{$\downarrow$} & \multicolumn{1}{c}{$\downarrow$} & \multicolumn{1}{c}{$\uparrow$} & \multicolumn{1}{c}{$\uparrow$} & \multicolumn{1}{c}{Replication} & \multicolumn{1}{c}{Audio, Video} \vspace{0.1cm} \\ 
		Ma et al.~\cite{ma2018evaluation}  & \multicolumn{1}{c}{$\uparrow$} & \multicolumn{1}{c}{$\uparrow$} & \multicolumn{1}{c}{$\uparrow$} & \multicolumn{1}{c}{$\uparrow$} & \multicolumn{1}{c}{$\uparrow$} & \multicolumn{1}{c}{$\uparrow$} & \multicolumn{1}{c}{Timing for TDM} & \multicolumn{1}{c}{Audio, Video}  \vspace{0.1cm} \\   
		Mahdian et al.~\cite{mahdian2018mindelay}  	& \multicolumn{1}{c}{$\downarrow$} & \multicolumn{1}{c}{$\uparrow$} & \multicolumn{1}{c}{$\downarrow$} & \multicolumn{1}{c}{$\downarrow$} & \multicolumn{1}{c}{$\uparrow$} & \multicolumn{1}{c}{$\uparrow$} & \multicolumn{1}{c}{Caching} & \multicolumn{1}{c}{Audio, Video} \vspace{0.1cm} \\ 
		Merlin et al.~\cite{merlin2018latency} 	& \multicolumn{1}{c}{$\uparrow$} & \multicolumn{1}{c}{$\downarrow$} & \multicolumn{1}{c}{$\uparrow$} & \multicolumn{1}{c}{$\downarrow$} & \multicolumn{1}{c}{$\uparrow$} & \multicolumn{1}{c}{$\uparrow$} & \multicolumn{1}{c}{Latency aware} & \multicolumn{1}{c}{Audio, Video} \vspace{0.1cm} \\ 
		Suksomboon et al.~\cite{suksomboon2018configuring}	& \multicolumn{1}{c}{$\downarrow$} & \multicolumn{1}{c}{$\uparrow$} & \multicolumn{1}{c}{$\uparrow$} & \multicolumn{1}{c}{$\uparrow$} & \multicolumn{1}{c}{$\uparrow$} & \multicolumn{1}{c}{$\downarrow$} & \multicolumn{1}{c}{Predic. Based} & \multicolumn{1}{c}{Audio, Video} \vspace{0.1cm} \\ 
		Ngo et al.~\cite{ngo2018efficient}	& \multicolumn{1}{c}{$\uparrow$} & \multicolumn{1}{c}{$\uparrow$} & \multicolumn{1}{c}{$\downarrow$} & \multicolumn{1}{c}{$\uparrow$} & \multicolumn{1}{c}{$\downarrow$} & \multicolumn{1}{c}{$\downarrow$} & \multicolumn{1}{c}{Estimation} & \multicolumn{1}{c}{DDoS Protec.} \vspace{0.1cm} \\ 
		Kim et al.~\cite{kim2018latency} & \multicolumn{1}{c}{$\downarrow$} & \multicolumn{1}{c}{$\downarrow$} & \multicolumn{1}{c}{$\uparrow$} & \multicolumn{1}{c}{$\downarrow$} & \multicolumn{1}{c}{$\uparrow$} & \multicolumn{1}{c}{$\downarrow$} & \multicolumn{1}{c}{Graph maint.} & \multicolumn{1}{c}{NFVs} \vspace{0.1cm} \\  \hline 
		\multicolumn{8}{c}{\textbf{Routing}} \\
		Koutsiamanis et al.~\cite{koutsiamanis2018best} 	& \multicolumn{1}{c}{$\uparrow$} & \multicolumn{1}{c}{$\uparrow$} & \multicolumn{1}{c}{$\uparrow$} & \multicolumn{1}{c}{$\downarrow$} & \multicolumn{1}{c}{$\uparrow$} & \multicolumn{1}{c}{$\downarrow$} & \multicolumn{1}{c}{Data dupl.} & \multicolumn{1}{c}{Wireless} \vspace{0.1cm} \\ 
		Levy et al.~\cite{levy2019establishing} & \multicolumn{1}{c}{$\uparrow$} & \multicolumn{1}{c}{$\uparrow$} & \multicolumn{1}{c}{$\uparrow$} & \multicolumn{1}{c}{$\uparrow$} & \multicolumn{1}{c}{$\downarrow$} & \multicolumn{1}{c}{$\uparrow$} & \multicolumn{1}{c}{Tree-Leaf Forw.} & \multicolumn{1}{c}{Multicast} \vspace{0.1cm} \\ 
		Khan~ et al.~\cite{khan2018traffic} & \multicolumn{1}{c}{$\uparrow$} & \multicolumn{1}{c}{$\downarrow$} & \multicolumn{1}{c}{$\downarrow$} & \multicolumn{1}{c}{$\downarrow$} & \multicolumn{1}{c}{$\downarrow$} & \multicolumn{1}{c}{$\uparrow$} & \multicolumn{1}{c}{Traffic-Aware} & \multicolumn{1}{c}{Vehicular} \vspace{0.1cm} \\ 
		Thubert et al.~\cite{thubert2019deterministic} 	& \multicolumn{1}{c}{$\downarrow$} & \multicolumn{1}{c}{$\uparrow$} & \multicolumn{1}{c}{$\uparrow$} & \multicolumn{1}{c}{$\uparrow$} & \multicolumn{1}{c}{$\uparrow$} & \multicolumn{1}{c}{$\uparrow$} & \multicolumn{1}{c}{Seg. \& Domain Info.} & \multicolumn{1}{c}{Audio, Video} \vspace{0.1cm} \\ 
		Kim et al.~\cite{kim2018kreonet} & \multicolumn{1}{c}{$\uparrow$} & \multicolumn{1}{c}{$\uparrow$} & \multicolumn{1}{c}{$\uparrow$} & \multicolumn{1}{c}{$\uparrow$} & \multicolumn{1}{c}{$\downarrow$} & \multicolumn{1}{c}{$\uparrow$} & \multicolumn{1}{c}{SDN} & \multicolumn{1}{c}{Wide Area Net.} \vspace{0.1cm} \\ 
		Grant et al.~\cite{grant2018new} & \multicolumn{1}{c}{$\uparrow$} & \multicolumn{1}{c}{$\downarrow$} & \multicolumn{1}{c}{$\downarrow$} & \multicolumn{1}{c}{$\downarrow$} & \multicolumn{1}{c}{$\downarrow$} & \multicolumn{1}{c}{$\uparrow$} & \multicolumn{1}{c}{Traffic Info.} & \multicolumn{1}{c}{5G Appl.} \vspace{0.1cm} \\ 
		Chen et al.~\cite{chen2019etn} 	& \multicolumn{1}{c}{$\uparrow$} & \multicolumn{1}{c}{$\uparrow$} & \multicolumn{1}{c}{$\uparrow$} & \multicolumn{1}{c}{$\uparrow$} & \multicolumn{1}{c}{$\downarrow$} & \multicolumn{1}{c}{$\uparrow$} & \multicolumn{1}{c}{Eth. Design} & \multicolumn{1}{c}{5G, Metro, DC} \vspace{0.1cm} \\ 
		Pointurier et al.~\cite{pointurier2019end} 	& \multicolumn{1}{c}{$\uparrow$} & \multicolumn{1}{c}{$\uparrow$} & \multicolumn{1}{c}{$\uparrow$} & \multicolumn{1}{c}{$\uparrow$} & \multicolumn{1}{c}{$\uparrow$} & \multicolumn{1}{c}{$\uparrow$} & \multicolumn{1}{c}{Timing} & \multicolumn{1}{c}{Industry~4.0} \vspace{0.1cm} \\ 
		Bocquillon~\cite{bocquillon2018robust} & \multicolumn{1}{c}{$\uparrow$} & \multicolumn{1}{c}{$\downarrow$} & \multicolumn{1}{c}{$\downarrow$} & \multicolumn{1}{c}{$\downarrow$} & \multicolumn{1}{c}{$\uparrow$} & \multicolumn{1}{c}{$\downarrow$} & \multicolumn{1}{c}{Optimization} & \multicolumn{1}{c}{Reliable Net.} \vspace{0.1cm} \\ 
		Borah et al.~\cite{borah2018energy} & \multicolumn{1}{c}{$\downarrow$} & \multicolumn{1}{c}{$\uparrow$} & \multicolumn{1}{c}{$\uparrow$} & \multicolumn{1}{c}{$\uparrow$} & \multicolumn{1}{c}{$\uparrow$} & \multicolumn{1}{c}{$\downarrow$} & \multicolumn{1}{c}{Predication Based} & \multicolumn{1}{c}{Energy conser.} \vspace{0.1cm} \\ 
		Jalan et al.~\cite{jalan2018forwarding} & \multicolumn{1}{c}{$\uparrow$} & \multicolumn{1}{c}{$\uparrow$} & \multicolumn{1}{c}{$\uparrow$} & \multicolumn{1}{c}{$\uparrow$} & \multicolumn{1}{c}{$\uparrow$} & \multicolumn{1}{c}{$\uparrow$} & \multicolumn{1}{c}{Service Policy} & \multicolumn{1}{c}{Audio, Video} \vspace{0.1cm} \\ 
		Thubert et al.~\cite{thubert2019transport} & \multicolumn{1}{c}{$\downarrow$} & \multicolumn{1}{c}{$\downarrow$} & \multicolumn{1}{c}{$\downarrow$} & \multicolumn{1}{c}{$\downarrow$} & \multicolumn{1}{c}{$\uparrow$} & \multicolumn{1}{c}{$\uparrow$} & \multicolumn{1}{c}{Trans. order} & \multicolumn{1}{c}{Audio, Video}  \vspace{0.1cm} \\   
		Thubert et al.~\cite{thubert2018bit} & \multicolumn{1}{c}{$\uparrow$} & \multicolumn{1}{c}{$\uparrow$} & \multicolumn{1}{c}{$\uparrow$} & \multicolumn{1}{c}{$\downarrow$} & \multicolumn{1}{c}{$\uparrow$} & \multicolumn{1}{c}{$\downarrow$} & \multicolumn{1}{c}{Replication} & \multicolumn{1}{c}{Audio, Video}  \vspace{0.1cm} \\   
		Wetterwald et al.~\cite{wetterwald2018insertion} & \multicolumn{1}{c}{$\uparrow$} & \multicolumn{1}{c}{$\uparrow$} & \multicolumn{1}{c}{$\uparrow$} & \multicolumn{1}{c}{$\downarrow$} & \multicolumn{1}{c}{$\uparrow$} & \multicolumn{1}{c}{$\downarrow$} & \multicolumn{1}{c}{Slot Reserv.} & \multicolumn{1}{c}{Mobile Nodes}  \vspace{0.1cm} \\    \hline  
		\multicolumn{8}{c}{\textbf{Segment Routing (SR)}} \\
		Bashandy et al.~\cite{bashandy2019segment} & \multicolumn{1}{c}{$\downarrow$} & \multicolumn{1}{c}{$\uparrow$} & \multicolumn{1}{c}{$\uparrow$} & \multicolumn{1}{c}{$\downarrow$} & \multicolumn{1}{c}{$\uparrow$} & \multicolumn{1}{c}{$\uparrow$} & \multicolumn{1}{c}{LDP} & \multicolumn{1}{c}{Audio, Video}  \vspace{0.1cm} \\   
		Filsfils et al.~\cite{filsfils2019scalable} & \multicolumn{1}{c}{$\downarrow$} & \multicolumn{1}{c}{$\downarrow$} & \multicolumn{1}{c}{$\downarrow$} & \multicolumn{1}{c}{$\uparrow$} & \multicolumn{1}{c}{$\downarrow$} & \multicolumn{1}{c}{$\uparrow$} & \multicolumn{1}{c}{Sub-path Trees} & \multicolumn{1}{c}{Audio, Video}  \vspace{0.1cm} \\   
		Laberge et al.~\cite{laberge2018tactical} & \multicolumn{1}{c}{$\downarrow$} & \multicolumn{1}{c}{$\downarrow$} & \multicolumn{1}{c}{$\uparrow$} & \multicolumn{1}{c}{$\downarrow$} & \multicolumn{1}{c}{$\uparrow$} &\multicolumn{1}{c}{$\uparrow$} &  \multicolumn{1}{c}{Optimization} & \multicolumn{1}{c}{Audio, Video}  \vspace{0.1cm} \\   
		Katsalis et al.~\cite{katsalis2018towards} & \multicolumn{1}{c}{$\uparrow$} & \multicolumn{1}{c}{$\uparrow$} & \multicolumn{1}{c}{$\uparrow$} & \multicolumn{1}{c}{$\uparrow$} & \multicolumn{1}{c}{$\downarrow$} & \multicolumn{1}{c}{$\uparrow$} & \multicolumn{1}{c}{Eth. Design} & \multicolumn{1}{c}{VPN, VLAN}  \vspace{0.1cm} \\   
		Chunduri et al.~\cite{chu2018pre,eckert2018preferred} & \multicolumn{1}{c}{$\downarrow$} & \multicolumn{1}{c}{$\uparrow$} & \multicolumn{1}{c}{$\uparrow$} & \multicolumn{1}{c}{$\uparrow$} & \multicolumn{1}{c}{$\uparrow$} & \multicolumn{1}{c}{$\downarrow$} & \multicolumn{1}{c}{Per hop proc.} & \multicolumn{1}{c}{Audio, Video}  \vspace{0.1cm} \\   
		Wang et al.~\cite{wang2019sdn} & \multicolumn{1}{c}{$\uparrow$} & \multicolumn{1}{c}{$\uparrow$} & \multicolumn{1}{c}{$\uparrow$} & \multicolumn{1}{c}{$\downarrow$} & \multicolumn{1}{c}{$\uparrow$} & \multicolumn{1}{c}{$\uparrow$} & \multicolumn{1}{c}{Inter-DC} & \multicolumn{1}{c}{Multipath Traff.}  \vspace{0.1cm} \\   \hline  
	\end{tabular}
\end{table*}

\section{Research Efforts: Traffic Shapers, Queuing, and Routing for Long-Distance Low-Latency Communications} \label{res:sec}
\subsection{Traffic Shapers}

\subsubsection{TAS Refinements: Adaptive Bandwidth Sharing (ABS) and Adaptive Slot Windows (ASW)} \label{ABSASW:sec}

The IEEE 802.1 Qbv TAS has recently been
refined with an ABS mechanism and an ASW
mechanism~\cite{nas2019per}. The ABS mechanism dynamically shares the
bandwidth of the respective ST and BE windows when the corresponding
traffic class has no traffic to send and would let the bandwidth go
unused. For instance, when all queued ST traffic has been transmitted,
but there is still time left in the ST window, then the ABS mechanism
transmits BE traffic (if there is queued BE traffic) in the remainder
of the ST window. Similarly, the ABS mechanism transmits ST traffic in
a BE window if all queued BE traffic has been transmitted and there is
time left in the BE window. The ABS mechanism is a simple low
complexity refinement to the TAS shaper that can reduce delays while
still enforcing the regular TAS timing guarantees.

The ASW mechanism feeds back ST traffic delay measurements from the
sink nodes upstream. Based on these ST traffic delays, the upstream
switches adjust their ST to BE traffic gating ratios to keep the ST
delay within a desired range. The ASW mechanism adds some complexity,
mainly due to the upstream signalling of the measured ST traffic
delays. A key advantage of the ASW mechanism is that it can
accommodate variations of the ST vs. BE traffic composition,
independent of the initial setting of the ST to BE gating ratio when
the network is initialized.

The standard TAS has been mainly designed for small scale networks.
The ABS and ASW refinements make TAS more flexible. It is an open
question whether these expanded flexibilities are sufficient to make
TAS suitable for LDNs. Possibly, a registration and reservation
protocol needs to reconfigure the network by scaling up the cycle time
to be on the order of the scaled up switch-to-switch propagation
delays in LDNs while considering the traffic volumes of the flows on
each switch port.

\subsubsection{Paternoster Policing and Scheduling}
The Paternoster algorithm is a proposed enhancement by Mike
Seaman~\cite{seaman2019pat} to standard CQF.  Paternoster provides
bounded latencies and lossless service for flows that are successfully
registered across the network without a time synchronization
requirement. For each egress port, the Paternoster protocol defines a
counter for stream reservation and four output queues (\textit{prior},
\textit{current}, \textit{next}, \textit{last}), whereby all switches
under Paternoster operate under an \textit{epoch} timescale which are
not synchronized with each other. In each \textit{epoch} window,
frames in the \textit{prior} queue are transmitted first until all
frames are transmitted. Once the \textit{prior} queue is depleted, the
\textit{current} queue is selected for transmission until the end of
the current epoch. While frames are being transmitted from the
\textit{prior} and \textit{current} queues, received frames are
enqueued in the \textit{current} queue until the bandwidth capacity is
reached for the current \textit{epoch}. Any additional frames are
enqueued in the \textit{next} and \textit{last} queues in a similar
manner, i.e., until the reservation capacity for the current epoch is
reached while additional frames are dropped if the \textit{last} queue
is completely reserved for the current \textit{epoch}. Note that all
ST traffic streams are given guaranteed bandwidth, while BE traffic is
given the leftover bandwidth. When a new epoch starts, the previous
\textit{current} queue operates as the \textit{prior} queue while the
\textit{next} and \textit{last} queues become the \textit{current} and
\textit{next} queues, respectively. The previous \textit{prior} queue
(which should be empty) becomes the new \textit{last} queue. The
Paternoster operation repeats at each \textit{epoch}, while the four
queues alternate during each \textit{epoch}.  While four queues are
expected to be sufficient for many LDN scenarios, very long
propagation delays may necessitate that another queue into the past
and another queue into the future are added, for a total of six
queues~\cite{seaman2019pat}.

Zhou et al.~\cite{zhouinsight,zho2018ana} have conducted a simulation
study on Paternoster, but only for one-hop transmission (they did not
consider a full multi-hop network) .

In summary, the Paternoster approach uses four queues that alternate
every epoch (or cycle) using only frequency synchronization, i.e., the
epoch duration is the same across the nodes. In contrast to CQF, the
Paternoster approach gives up some delay predictability in exchange
for not requiring clock synchronization and for reducing the average
delay.  There has only been one limited Paternoster comparison study
by Zhou et al., 2018 which considered only one-hop transmission, not a
full multi-hop network, and compared Paternoster with synchronized
scheduling, namely 802.1Qbv TAS and 802.1Qch CQF with two queues.  A
comprehensive study of Paternoster in comparison to the other
approaches, which can operate without time synchronization, i.e., SDF
and CSQF, across a wide set of operating conditions is needed; and in
comparison to the approaches requiring time synchronization.

\subsubsection{Other Approaches}
One of the applications that require low latencies
over long distances are the smart-grid. Additionally, smart grids are
critical infrastructure the require high reliability. Hence,
Ball et. al~\cite{ball2019evaluation} have presented smart grid synchrophaser
measurements and a control systems design
over a wide area network. Their main
design focus is to ensure real-time communication requirements over
long distances. To achieve this, the authors propose to incorporate
strict priority queuing, static routing rules for time critical
traffic, redundant transmissions over error recovery,
limiting the forwarding rule lookup to only header based lookup,
and a predictable traffic knowledge for path evaluations.
In order to evaluate deterministic latencies the system uses a
fixed number of bits to forward between nodes, and compares against a
universal time to compute the time elapsed between end nodes, and then
uses this information to synchronize the nodes to forward the data
required to maintain the control loop.

The scheduling of traffic has been discussed in
Specht et al.~\cite{Specht2016}, where the urgency based scheduler
to forward the time sensitive traffic based on priorities has been presented.
The buffer management for strict priority scheduling has been discussed in
Mangin et al.~\cite{mangin2019multiplexing}, where
memory buffers are associated with a dedicated priority and
queue elements are then sorted based on the priority and
timestamps within a given buffer before the selection for
the transmissions.

Li et al.~\cite{li2019enhanced} presented a technique for IoT networks
to provide time sensitive properties such as real time for
end-devices through time-triggered networks.
The main concern for time-sensitive applications
is to preserve the deterministic properties in the midst of
reconfiguration and changing network conditions in the IoT.
SDN principles that are applied in wide area networks,
and data center networks cannot be directly applied to
IoT networks since the time sensitive properties are
not embedded in both control and data plane operations.
As a result, the IoT networks must consider deterministic
transmission in both control and data planes by prioritizing
traffic across the network. To establish time-sensitive connections in wireless networks,
Buratti~\cite{buratti2018joint} proposed a method in which
the destination is responsible for understanding the network
topology to assign the time sensitive shaping properties
to the forwarding nodes. Whereas, Said et al.~\cite{said2019sdn}
provide a mechanism to maintain and update the TSN
configurations of traffic shaping for new devices that are to
be integrated into an ongoing TSN network. More specifically, the
central focus of this article is to reduce the
time-to-integrate delay when a new device is introduced into
the network by exploiting the IEEE 802.1Qcc model and IEEE 802.1AS
in the context of SDN centralized configuration mechanisms.
The time-to-integrate time is a necessary
factor for applications that include a frequent introduction
of a new sensors on to an existing sensor grid infrastructure,
especially in the industrial and  automotive contexts, where
a new communicating sensor must be tested and added
to the main network to balance the load.
Wetterwald~\cite{wetterwald2018source} have proposed a mechanism
to add the deterministic schedules on the network path
between source and the destination by considering the
start time at the source and the source-routed mechanism. To
ensure this, the source node has the overview of entire network
and a configuring agent creates
and forwards the schedule over a deterministic network
to establish a deterministic path.

\subsection{Queuing and Forwarding}
Queuing of packets on the forwarding nodes determines the overall time
of the packet spent waiting on the node. Therefore, it is important to
consider queuing policies carefully when designing time-critical
networks. Flow based schedulers proposed by
Joung et al.~\cite{joung2019regulating} in traditional integrated
services (IntServ) framework
have complexities of $O(N)$ or $O(\log
N)$, where $N$ is the number of flows in the scheduler, which can grow to
tens of thousands in a core router. Due to such complexity,
class-based schedulers are typically adopted in real deployments. The
class-based systems, however, cannot provide bounded delays in cyclic
networks, since the maximum burst grows infinitely along
the cycle path.
Therefore, Joung et al.~\cite{joung2019regulating}
consider a conserving fair schedulers knows as the
Regulating Schedulers (RSC). RSC acts as both as a
regulator and a scheduler to achieve the fairness in the
scheduling. A deficit round-robin
(DRR) based RSC provides both
regulating and scheduling functions for a given port.
In addition to lower complexity, the input port-based DRR
is shown that the forwarding process is between than
TSN approach. DRR can satisfy end-to-end latency
bound on the order of milliseconds for realistic
network scenarios. Whereas, for a Delay Tolerant Network (DTN),
Ayub et al.~\cite{ayub2018priority}
have presented a mechanism to address the congestion originating
from multiple copies (replicas) of packets that are sent for reliability.
Multi-copy routing protocols duplicates the packets
which results in a network congestion. In order to
avoid the congestion, network could drop the packets that
are being process. The dropping of packets should be done
in a controlled way such that there is no negative impacts
to the reliability mechanism. Ayub et al.~\cite{ayub2018priority}
mainly considers reactive dropping, i.e., dropping of packets
only occurs when the queue overflows. This is achieved
by a Priority Queue Based Reactive Buffer Management
Policy (PQB-R) in an urban environment scenario. The PQB-R
mainly categorizes the enqueued packets into
three different queues and enforces a separate drop metric on
each queue, thus creating an class based dropping mechanism.
The experimental results presented in the article
demonstrate that the proposed PQB-R has
reduced overall low number of packets in the network due to packet
drops which results in an increased delivery ratio.
As an alternative or complement to multiple packet copies, future work
may explore the use of low-latency networking coding mechanisms to
improve the reliability while keeping latencies
low~\cite{dou2017ins,gab2018cat,tou2011fly,wun2017cat}.

In an effort to understand the detailed characteristics
of a scheduling and queuing model,  Ma et al.~\cite{ma2018evaluation}
proposed a scheduling model for a
Flexilink which is a newly proposed dynamic TDM network protocol and
architecture that strives to be secure and stable. The proposed
delay-based Flexilink approach is compared with classic best effort
and priority based scheduling through simulations. The results
indicate that the proposed scheduling algorithm performs better, even
when the network is heavily loaded.

Managing the congestion while devising a queuing mechanism is an
important effort to preserve time-sensitive properties of the network.
Towards this end, the study by Mahdian et al.~\cite{mahdian2018mindelay}
has presented a framework for caching networks to jointly optimize
forwarding and caching strategies for minimizing congestion-dependent
network cost. Caching variables are typically integer valued which
results in an NP-hard optimization problem.
Hence to reduce the optimization complexity, authors
propose a technique where caching variables are extended to be
real-valued which reduces the optimization complexity.
Authors also present the optimality conditions necessary for
the real-valued optimization problem.
The proposal is then extended to devise an
adaptive and distributed joint forwarding and caching algorithm,
MinDelay. MinDelay optimization is based on a conditional
gradient approach which can be implemented in a distributed
manner. MinDelay approach also results in a low complexity
and overhead for caching and forwarding mechanism
Evaluation results for MinDelay show
significantly better delay performance in the
low to moderate request rate regions over a wide range of network
topologies.
The follow-up study to MinDelay
by Melin et al.~\cite{merlin2018latency} introduced
Latency-Aware Forwarding for an Intrinsically Resilient Overlay Network
(referred to as IRON). IRON is based on Back-Pressure Forwarding (BPF)
and supports latency-sensitive traffic. Latency-Aware Forwarding adds
support for latency-sensitive traffic while maintaining the BPF
throughput optimality for latency-insensitive flows. Latency-Aware
Forwarding combines a number of advances to a) forward
latency-constrained packets along delay-appropriate paths and to b)
reduce the processing time of these packets at each hop. The
evaluations in \cite{merlin2018latency} compare Latency-Aware
Forwarding to traditional and work-conserving BPF and indicate a 233\%
increase in goodput in delivery of delay-constrained traffic.

To ensure the minimum queuing impact from the network
configuration process,  Suksomboon et al.~\cite{suksomboon2018configuring}
have proposed a performance characterization of a software router by
conducting a packet latency prediction model based on the
Erlang-k distribution.
The prediction model designed requires only limited observation
from the queues of the network interface card, assuming that
traffic belonging to multiple configurations arrive at a
port over to a common queue. The average latencies are then estimated
for each configuration on the network.
The estimation of latencies by the prediction model also helps
in the configuration selection (CS) such that the configuration
that results in the minimum average packet latency can be chosen
for an application.

The queuing and forwarding mechanism can be compromised through
attacks from rouge entities. As one of the main types of
Distributed Denial of Service (DDoS) attacks is the SYN flood
attack, which results in service denial for legitimate clients.
This occurs due to the overwhelming service requests to service
by the attacker. The article Ngo et al.~\cite{ngo2018efficient}
introduces an efficient high-throughput, and
low-latency SYN flood defender architecture.  SYN flood is
devised through a mathematical modeling in which
the estimation architecture identifies SYN flood attacks in both
throughput and latency. A novel prototype based on Verilog-HDL modeling
has been evaluated for the high-rate SYN flood attacks,
which can be integrated into an OpenFlow switch for handling network
packets. he evaluations with NetFPGA-10G platforms showed that
the core can protect servers against SYN flood attacks for nearly
more than 28 millions packets per second which is significantly
better than traditional hardware-based approaches.

As Network Function Virtualization (NFV) and
Software Defined Networking (SDN) technologies gradually mature as
next-generation network technologies, management and orchestration
(MANO) technologies that manage the Service Function Chain (SFC) have
received extensive research attention~\cite{cao2019dyn,sha2018lay,xia2019red,yur2018ope}.
Kim et al.~\cite{kim2018latency} have proposed a Graph Selection Manager
(GSM) to provide one or more VNF forwarding graphs given a
maximum latency bound as well as VNFs and network capacity.
The emulation evaluations indicate that one
or more VNF forwarding graphs can meet the service level agreement
(SLA) of a tenant in scenarios with limited network capacity and can
establish multiple end-to-end low-latency network services.

\subsection{Routing}
Traditionally, routing of a flow through multiple paths has relied on
simple lookup of the next path based on limited information of source
and destination nodes which ignores the time-sensitive properties of
delivering packet between end-points~\cite{bar2014nov,guc2017uni}. Therefore, for long distance
communications, routing has to consider and incorporate time-sensitive
properties in the path determination process.  For instance, the IEEE
802.15.4 Time-Slotted Channel Hopping (TSCH) medium access control
mechanism uses traditional collision detection and retransmission
procedures that cannot enforce the end-to-end time-sensitive
communications.  Therefore, Koutsiamanis et
al.~\cite{koutsiamanis2018best} have propose to use LeapFrog
Collaboration (LFC) on top of a Routing Protocol (RPL) for
establishing deterministic and reliable communication between
end-points.  The LFC algorithm duplicates the data flow onto an
alternate path with a goal to exploit route diversity to achieve low
latency and reliability. In another effort, RFC 2210~\cite{rfc2210}
provides a route reservation protocol to for integrated
services. Routing multicast traffic in a time-sensitive environment
requires the synchronization of time-sensitive configuration across
multiple nodes over multiple path which need to be simultaneously
configured.  Levy et al.~\cite{levy2019establishing} have proposed a
multicast forwarding tree that originates from a root where a single
multicast source, as a root, forwards configuration information to a
set of leaf nodes to configure the leaf nodes such that a
multipath flow arrives simultaneously at the terminal destination
nodes.

One of the standard applications of TSN networks are the vehicular
networks. Vehicular ad-hoc networks (VANETs) require low-delay routes for time-critical traffic associated with the sensor
and control systems that reside in cars.  Khan~ et
al.~\cite{khan2018traffic} proposed a Traffic Aware Segment-based
Routing (TASR) protocol which considers an Expected Connectivity
Degree (ECD) that includes the vehicle density information, and
geographical information of different segments between source and
destination nodes to evaluate the routing path. In an effort to
maintain the scalability of deterministic flows over the forwarding
nodes, Thubert et al.~\cite{thubert2019deterministic} has presented a
method for categorizing the deterministic networks based on
deterministic segments and deterministic domains.  The resources are
then allocated to the deterministic segments and domains based on the
flows that are supported on segments and domains.  Similarly for long
distance communications, Kim et al.~\cite{kim2018kreonet} have
presented a large scale infrastructure, KREONET-S, designed as a
Software Defined Wide Area Network (SD-WAN) in Korea focusing on
delivering time critical end-to-end connectivity in WAN networks. The
results from their deployment showed improved network throughput,
minimal delay, and constant jitter which are necessary to host the
time-sensitive applications over a WAN network.

In contrast to SDN, a distributed mechanism for establishing an
end-to-end routing path requires coordination among forwarding nodes.
In conjunction to SD-WAN networks, 5G networks provide long range low
latency solutions. Within the context of 5G, the time-sensitive
applications over IP networking has been discussed in Grant et
al.~\cite{grant2018new}. In support of deterministic forwarding
latencies over 5G networks, Chen et al.~\cite{chen2019etn} have
presented an Ethernet design which can support the requirements for 5G
mobile transport, metro, and data center interconnects networks. The
Ethernet design comprehensively supports multi-service access,
deterministic forwarding latency, hard traffic isolation, hierarchical
traffic multiplexing, and flexible forwarding across L1, L2 and L3
networks, and multi-layer Operations and Management (OAM) mechanism.
Time-sensitive applications in 5G networks include fronthaul and
Industry 4.0 which require strict deterministic requirements with zero
jitter. Pointurier et al.~\cite{pointurier2019end} have review and
discuss the current solutions, as well as present research directions
to support diverse set of applications that require time-sensitive
properties within the network.

Typically, forwarding nodes communicate the routing information when
there is a change in the network, such as the introduction of a new
node or a node failure. The traffic related to the coordination of a
routing path during a network change has to be robust to ensure
minimum disruption to the
network. Bocquillon~\cite{bocquillon2018robust} proposed a system to
provide a delay-/disruption-tolerant network (DTN). More specifically,
this article proposes an algorithm as a robust mechanism that
minimizes the dissemination length of the messages that needs to be
transferred between source and forwarding nodes in deciding the
routing plan.

One of the downsides of reliability from packet replication is that
replication increases the energy consumption. In an effort to ensure
reliability and to conserve overall energy spent for ensuring the
reliability, Borah et al.~\cite{borah2018energy} have presented a
energy-ware routing protocol, Energy-efficient Location
Prediction-based Forwarding for Routing using Markov Chain
(ELPFR-MC). ELPFR-MC has been developed for opportunistic networks
(OppNets). However, OppNets are a subclass of delay-tolerant networks,
which can be extended to adapt for time-sensitive networks to conserve
both time-sensitive properties and energy efficiency.

Jalan et al.~\cite{jalan2018forwarding} presented a mechanism in which
packets are forwarded within the nodes based on a service policy. That
is, when a gateway node receives a packet, the packet is matched with
an on-going service policy of the network, and the forwarding path is
based on service policies supported on the nodes. A service address is
used to identify the service data, and service policy of the
network. New services can be added to the network through service
configurations. The service based approach reduces the complexity of
network to establish routes according on a broader service policy
based allocation of resources as opposed to flow based requirements.

The storing and forwarding of packets through the transport layer of a
device typically does not consider the application information, such
as their priorities and flow properties.  To address such a forwarding
mechanism, Thubert et al.~\cite{thubert2019transport} have provided a
method to track and insert identifiers such that the receiving node
can order and package the incoming flow in the order in which it was
transmitted. If the links are deterministic, the flow of traffic over
multiple nodes where the packet flow order is preserved can ensure the
end-to-end connection to have deterministic properties.

In an effort to ensure the transmission redundancy for the required
QoS between the end points, Thubert et al.~\cite{thubert2018bit} have
described a method in which, for each packet there is a bit index,
such that each bit in the index maps to a deterministic segment. When
the packet traverses through the network segments, the corresponding
bit within the bit index of the packet is used to decide the
replication process for that packet. Thus, a source can control the
replication factors by setting and un-setting the bit index fields to
ensure the end-to-end QoS needed between the source and destination.

Establishing a deterministic end-to-end link could be particularly
hard in nodes that are not stationary.  If the intermediate
forwarding nodes are mobile and moving frequently, one solution for
the end-to-end flow establishment is to use a centralized
configuration, such as through SDN. The downsides of centralized
configuration are the control plane latency, overhead, and computation
requirements.  Alternative to centralized configuration, Thubert et
al.~\cite{thubert2019insertion} have presented an interesting approach to
reserve a slot at each forwarding node for an end-to-end flow once
established. Such that, if a node changes its location, and when the
packet arrives on a deterministic path, then the resources for that flow
would be reserved for processing and forwarding in the form of slots
which are inserted when the flow is established. This approach assumes
that there exists a deterministic path for the packets to arrive at the end
point considering that packet flows through source and intermediate
nodes that are mobile, but same number of nodes that ordered in
different way. Suppose A, B, C, and D are the nodes in the
flow path, whereby A and D are the end-points. The positions of B and C
could be interchanged with the insertion slot method, whereby A could
forward the packet to either B or C while effectively preserving the
QoS properties. This technique can be extended to large number of
nodes, supporting long distances. Similar to an insertion slot, Wetterwald
et al.~\cite{wetterwald2018insertion} have presented a slot reservation of
resources for end-to-end deterministic networking between
end-points.  Each flow is allocated with unique slots based on the QoS
requirements along the deterministic path between end points.

\subsection{Segment Routing (SR)}
Segment Routing (SR) aims to use MPLS (Multi-Protocol Label
Switching) and IPv6 segments to establish end-to-end connections with
deterministic properties.  SR policies perform traffic steering over
specific segments using segment identifiers which are configured as a
path.  SR could use SDN for interconnecting and configuring the
segment paths. Thus, with an extension of SR to a large number of
segments, an SR can be extended to achieve deterministic
characteristics over long distances which can provide multiple high
level Service Level Agreements (SLAs) over a given network between
end-points. However, one critical issue is to ensure the guaranteed
QoS over the SR network without compromising the flexibility and
scalability.

Bashandy et al.~\cite{bashandy2019segment} have described an SR method
using Label Distribution Protocol (LDP). LDP attaches a label to an
incoming packet with a segment ID.  The packet can then be forwarded
to another node over a Label Switched Path (LSP). Filsfils et
al.~\cite{filsfils2019scalable} extended their approach to address the
scalability and reliability specifically to improve the response time
to trace and correct a performance degradation in an SR path. The
system uses a Performance Measurement (PM) module to track the SR
segments for honoring the end-to-end SLAs over existing paths as well
as after addition of new segments and paths through the
re-configuration of the networks. PM allows the end-to-end flow
evaluation for delay-bound variations at sub-seconds level and
attempts to correct the network for any variation through
reconfiguration. These flow evaluations are then used to detect and
correct the SR configurations when a degradation occurs on the
end-to-end path.  For a given policy, PM can be difficult to achieve
for an SR when there are disjoint paths between end-points. Therefore,
an SR policy is divided into smaller sections that can be tracked as
part of Root-Nodes, and Sub-Path Trees (SPT). SPTs are used to track
the sub-paths associated with the root nodes, and root maintains
and run the PMs to track and action against again the performance
degrade. Filsfils et al.\cite{filsfils2019network} have also presented
a mechanism to implement the monitoring and end-to-end performance
evaluation through programmable functions for SR networks.  Similarly,
an efficient method for traffic monitoring through SR using a demand
matrix optimization framework has been presented by Laberge et
al.~\cite{laberge2018tactical}. The scalability of SR has also been
extensively studied in Jadin~\cite{jadin2019cg4sr}.  A column
generation method has been adapted to solve the large scale linear
programs pertaining to the long range SR segments.  There evaluations
show that near optimal solutions in creating SR paths can be achieved
that can also scale effectively for large topologies.

The advancement of SR which can enable dynamic path allocation with
deterministic properties has put pressure on the hardware
requirements. Traditional Ethernet networks are not designed to
track and adapt to the complex configurations required by SR.
For example, with the proliferation of virtualized network functions
in network deployments, the transport network between virtual
functions are increasingly dependent on Virtual Private Networks (VPN)
and Virtual Local Area Network (VLAN) for establishing end-to-end
connections. Katsalis et al.~\cite{katsalis2018towards} have identified that
the co-existence of VLAN and VPNs over a common infrastructure could
introduce a performance degradation due to congestion and competition over
common physical resources. Therefore, Katsalis et
al.~\cite{katsalis2018towards} have introduced a novel Ethernet design,
Flex-E, to support the resource slicing, flexibility, and scalability
for SR.

The protocols that support SR deployments are generally complex to
manage due to the increased reliance on protocol signalling
messages. As SR requires more configurations the underlying protocol
has to generate more signalling messages to configure the network
nodes. For example, the ReSource ReserVation Protocol with Traffic
Engineering (RSVP-TE) is typically used in SR applications, whereby
RSVP-TE generally limits the flexibility of dynamic scaling and
reconfiguration due to the computation and signalling requirements. In
an effort to reduce the signalling overhead of traditional protocols
that support SR, Chunduri et al.~\cite{chu2018pre,eckert2018preferred} have
proposed a Preferred Path Routing (PPR) protocol which signals the
routing information from computation engines to network nodes directly
through an existing message distribution, such as REST API's in SDN
network. The routing information includes explicit paths and per-hop
processing, e.g., QoS for deterministic forwarding.  PPR supports a
wide range of configuration capabilities including IP forwarding
planes and SR. As a result, PPR mechanisms result in a lightweight,
scalable, and flexible protocol to accommodate high precision network
services.  Chunduri et al.~\cite{chu2018pre,eckert2018preferred} have also presented an
enhancement to the PPR architecture, whereby PPR graphs are signaled to
forwarding nodes instead of point to point PPR paths to reduce the
overall signalling required to distribute the forwarding and QoS
entries across the network nodes. From such an approach, for
any-to-any connectivity of $N$ nodes, $O(N)$ scalability can be
achieved for PPR graphs forwarding entries.  In contrast, distributed
routing protocols, such as IGPs, incurred $O(N^2)$ complexity for
RSVP-TE with PPR point to point paths.
In summary, a main advantage of PPR over SR is that PPR reduces
the processing overhead of large headers in each node. Moreover, unlike SR,  PPR can pre-provision specific QoS parameters and
algorithms specific to a path based on the node capabilities in the network. This enables the seamless application of QoS algorithms to be enforced 
on the traffic with PPR-ID on the preferred path in the network.

Segment Routing can also be applied to Intra-data center (DC)
networks where the network can span long distances and require
deterministic characteristics for time sensitive applications such as,
tele-medicine hosted on multiple servers.  The server-to-server
communication traffic within a data center is characterized as
east-west traffic. In traditional inter-DC networks, the east-west
routing is managed by SDN, whereby, an Equal-cost multipath (ECMP) is
used for the traffic management. However, SDN management with ECMP
could be limited by an scalability issue arising from the limited
Ternary Content-Addressable Memory (TCAM) size in the forwarding
nodes. Wang et al.~\cite{wang2019sdn} have proposed an SDN-based
traffic engineering method, namely Dynamic-Flow-Entry Saving Multipath
(DFSM) for east-west traffic management to reduce the usage of TCAM
entries. Their evaluation results show that DFSM saves 15\% to 30\% of
TCAM flow entries over practical topologies, as well as reduces the
standard deviation of path latencies from 10\% to 7\% as compared to
label-switched tunneling which is typically used in SR.

\subsection{Discussion on Research Efforts of the Shapers Proposed for LDNs}
Long distance communication is an integral part in today's networking
applications and connectivity. However, one main challenge in long
distance communication is the resource management and the associated
delay for reconfigurations. Therefore, time sensitive applications
over long range communication should be designed carefully to ensure
that there are no side effects from the reconfiguration of the networks.

Applications that heavily rely on long distance
communication include smart grids, fronthaul telecommunications, and
dedicated point-to-point links, edge to Data Centers (DC) connectivity,
edge to user connectivity, and tiny DC deployments with
intra and inter connectivity requirements. Each application poses
a unique challenge to establish end-to-end deterministic characteristics. Although, research efforts are underway to address the challenges
in these areas, a careful consideration towards latency impact,
overhead, complexity, flexibility, cost, and dependency should be ensured.
\paragraph{Latency} The design should ensure that the latency is not
negatively impacted by a proposed mechanism while achieving a scalable
and flexible solution.
\paragraph{Overhead} The management traffic and the control plane data directly correspond to overhead. Distributed and centralized mechanisms both
have upsides and downsides in terms of re-configuration which have
independent implications for delay and the total required overhead. Further
research is required to find the balance between low overhead
and reconfiguration simplicity specifically for time-sensitive
long distance communications.
\paragraph{Complexity} Routing problems are generally solved on
a compute agent through an optimization framework. Complexities can
arise from the solution conversation, data reception, and data
dissemination to the actual forwarding nodes. Future research
should focus on solutions that in addition to
achieving near optimal solution, do not compromise the simplicity.
\paragraph{Flexibility and Scalability} A solution that attempts
to solve latency and overhead often does not consider the scalability;
often, the complexity increases exponential as the network grows. For
instance, in a distributed routing protocol the routing updates would
increase exponentially with the number of nodes. Thus, routing
protocol designs should carefully consider the flexibility and
scalability impact.
\paragraph{Cost} Cost is an important factor for the large scale
deployment and proliferation of a proposed technique to mainstream
networks.  For instance, although hardware based solutions provide
performance benefits, the cost and flexibility factors are
compromised.  Similarly, while software based solutions provide a cost
effective solution, latency and simplicity may be
compromised. Therefore, research efforts should focus on achieving a
balanced approach to keep these factors within a reasonable range, and
not to overshoot, while trying to optimize for a single factor.

\section{Conclusions and Future Work}  \label{sec:concl}
Overall, there has been extensive research and standardization towards
deterministic forwarding services using Ethernet technology. We have
presented a comprehensive survey on the recent advances in the
state-of-the-art TSN and DetNet forwarding protocols and have outlined
several limitations and advantages. Regrading the standards part, CQF
appears as the top choice in coordinating and ensuring TSN/DetNet
QoS. Several derivatives have been proposed (and are in draft
status). The existing proposed approaches in theory can be used to
efficiently and effectively provide deterministic QoS in large scale
networks according to several draft documents by Norman Finn without a
complete overhaul of the current network.
In terms of the quantitative efficacy of the approaches, it is
difficult to claim that CQF and its derivatives will work for all
cases since a pronounced lack of testing is apparent.

The research part for the deterministic forwarding shows that several
articles have addressed the DetNet QoS in large scale networks (e.g.,
Inter-DC, Mobile Backhaul, etc.). These studies indicate that
there is significant interest in pursuing deterministic behaviors in
LDNs.

In the future, an extensive evaluation of the main representatives of
the aforementioned state-of-the-art models is necessary. A rigorous
simulation study needs to be conducted to quantitatively examine the
efficacy of the CQF protocol and its main derivatives.

\bibliographystyle{IEEEtran}

\begin{thebibliography}{10}
\providecommand{\url}[1]{#1}
\csname url@samestyle\endcsname
\providecommand{\newblock}{\relax}
\providecommand{\bibinfo}[2]{#2}
\providecommand{\BIBentrySTDinterwordspacing}{\spaceskip=0pt\relax}
\providecommand{\BIBentryALTinterwordstretchfactor}{4}
\providecommand{\BIBentryALTinterwordspacing}{\spaceskip=\fontdimen2\font plus
\BIBentryALTinterwordstretchfactor\fontdimen3\font minus
  \fontdimen4\font\relax}
\providecommand{\BIBforeignlanguage}[2]{{%
\expandafter\ifx\csname l@#1\endcsname\relax
\typeout{** WARNING: IEEEtran.bst: No hyphenation pattern has been}%
\typeout{** loaded for the language `#1'. Using the pattern for}%
\typeout{** the default language instead.}%
\else
\language=\csname l@#1\endcsname
\fi
#2}}
\providecommand{\BIBdecl}{\relax}
\BIBdecl

\bibitem{finn2018introduction}
N.~Finn, ``Introduction to time-sensitive networking,'' \emph{IEEE
  Communications Standards Magazine}, vol.~2, no.~2, pp. 22--28, 2018.

\bibitem{messenger2018time}
J.~L. Messenger, ``Time-sensitive networking: An introduction,'' \emph{IEEE
  Communications Standards Magazine}, vol.~2, no.~2, pp. 29--33, 2018.

\bibitem{qiang-detnet-large-scale-detnet-04}
\BIBentryALTinterwordspacing
L.~Qiang, X.~Geng, B.~Liu, T.~Eckert, and L.~Geng, ``{Large-Scale Deterministic
  IP Network},'' Internet Engineering Task Force, Internet-Draft
  draft-qiang-detnet-large-scale-detnet-04, Mar. 2019, work in Progress.
  [Online]. Available:
  \url{https://datatracker.ietf.org/doc/html/draft-qiang-detnet-large-scale-detnet-04}
\BIBentrySTDinterwordspacing

\bibitem{IEEE8021Qat}
``{IEEE Standard for Local and Metropolitan Area Networks---Virtual Bridged
  Local Area Networks Amendment 14: Stream Reservation Protocol (SRP)},''
  \emph{IEEE Std 802.1Qat-2010 (Revision of IEEE Std 802.1Q-2005)}, pp. 1--119,
  Sep. 2010.

\bibitem{IEEE8021Qav}
``{IEEE Standard for Local and Metropolitan Area Networks - Virtual Bridged
  Local Area Networks Amendment 12 Forwarding and Queuing Enhancements for
  Time-Sensitive Streams},'' \emph{IEEE Std 802.1Qav-2009 (Amendment to IEEE
  Std 802.1Q-2005)}, pp. C1--72, Jan. 2009.

\bibitem{IEEE8021Qbu}
``{IEEE Standard for Local and metropolitan area networks -- Bridges and
  Bridged Networks -- Amendment 26: Frame Preemption},'' \emph{IEEE Std
  802.1Qbu-2016 (Amendment to IEEE Std 802.1Q-2014)}, pp. 1--52, Aug. 2016.

\bibitem{IEEE8023br}
``{IEEE Standard for Ethernet Amendment 5: Specification and Management
  Parameters for Interspersing Express Traffic},'' \emph{IEEE Std 802.3br-2016
  (Amendment to IEEE Std 802.3-2015 as amended by IEEE St802.3bw-2015, IEEE Std
  802.3by-2016, IEEE Std 802.3bq-2016, and IEEE Std 802.3bp-2016)}, pp. 1--58,
  Oct. 2016.

\bibitem{IEEE8021Qbv}
``{IEEE Standard for Local and metropolitan area networks -- Bridges and
  Bridged Networks - Amendment 25: Enhancements for Scheduled Traffic},''
  \emph{IEEE Std 802.1Qbv-2015 (Amendment to IEEE Std 802.1Q--- as amended by
  IEEE Std 802.1Qca-2015, IEEE Std 802.1Qcd-2015, and IEEE Std 802.1Q---/Cor
  1-2015)}, pp. 1--57, Mar. 2016.

\bibitem{IEEE8021Qci}
``{IEEE Standard for Local and metropolitan area networks--Bridges and Bridged
  Networks--Amendment 28: Per-Stream Filtering and Policing},'' \emph{IEEE Std
  802.1Qci-2017 (Amendment to IEEE Std 802.1Q-2014 as amended by IEEE Std
  802.1Qca-2015, IEEE Std 802.1Qcd-2015, IEEE Std 802.1Q-2014/Cor 1-2015, IEEE
  Std 802.1Qbv-2015, IEEE Std 802.1Qbu-2016, and IEEE Std 802.1Qbz-2016)}, pp.
  1--65, Sep. 2017.

\bibitem{IEEE8021Qch}
``{IEEE Standard for Local and metropolitan area networks--Bridges and Bridged
  Networks--Amendment 29: Cyclic Queuing and Forwarding},'' \emph{IEEE
  802.1Qch-2017 (Amendment to IEEE Std 802.1Q-2014 as amended by IEEE Std
  802.1Qca-2015, IEEE Std 802.1Qcd(TM)-2015, IEEE Std 802.1Q-2014/Cor 1-2015,
  IEEE Std 802.1Qbv-2015, IEEE Std 802.1Qbu-2016, IEEE Std 802.1Qbz-2016, and
  IEEE Std 802.1Qci-2017)}, pp. 1--30, Jun. 2017.

\bibitem{IEEE8021Qcc}
``{IEEE Draft Standard for Local and metropolitan area networks--Media Access
  Control (MAC) Bridges and Virtual Bridged Local Area Networks Amendment:
  Stream Reservation Protocol (SRP) Enhancements and Performance
  Improvements},'' \emph{IEEE P802.1Qcc/D2.0, October 2017}, pp. 1--207, Jan.
  2017.

\bibitem{nas2019ult}
A.~Nasrallah, A.~S. Thyagaturu, Z.~Alharbi, C.~Wang, X.~Shao, M.~Reisslein, and
  H.~ElBakoury, ``Ultra-low latency {(ULL)} networks: The {IEEE TSN} and {IETF
  DetNet} standards and related {5G ULL} research,'' \emph{IEEE Communications
  Surveys \& Tutorials}, vol.~21, no.~1, pp. 88--145, 2019.

\bibitem{IEEE8021Qcr}
J.~Specht, ``{IEEE Draft Standard for Local and metropolitan area
  networks--Media Access Control (MAC) Bridges and Virtual Bridged Local Area
  Networks Amendment: Asynchronous Traffic Shaping},'' \emph{IEEE
  P802.1Qcr/D0.4, April 2018}, Nov. 2017.

\bibitem{Specht2016}
J.~Specht and S.~Samii, ``Urgency-based scheduler for time-sensitive switched
  ethernet networks,'' in \emph{Proc. IEEE Euromicro Conf. on Real-Time
  Systems}, Jul. 2016, pp. 75--85.

\bibitem{nas2019per}
A.~Nasrallah, A.~S. Thyagaturu, Z.~Alharbi, C.~Wang, X.~Shao, M.~Reisslein, and
  H.~ElBakoury, ``Performance comparison of {IEEE 802.1 TSN Time Aware Shaper
  (TAS)} and {Asynchronous Traffic Shaper (ATS)},'' \emph{IEEE Access}, vol.~7,
  pp. 44\,165--44\,181, 2019.

\bibitem{bou2018the}
J.~{Le Boudec}, ``A theory of traffic regulators for deterministic networks
  with application to interleaved regulators,'' \emph{IEEE/ACM Transactions on
  Networking}, vol.~26, no.~6, pp. 2721--2733, Dec 2018.

\bibitem{mohammadpour2018end}
E.~Mohammadpour, E.~Stai, M.~Mohiuddin, and J.-Y.~L. Boudec, ``End-to-end
  latency and backlog bounds in time-sensitive networking with credit based
  shapers and asynchronous traffic shaping,'' \emph{arXiv preprint
  arXiv:1804.10608}, 2018.

\bibitem{moh2019imp}
E.~Mohammadpour, E.~Stai, and J.-Y.~L. Boudec, ``Improved delay bound for a
  service curve element with known transmission rate,'' \emph{arXiv preprint
  arXiv:1905.04956}, 2019.

\bibitem{ietf-detnet-architecture-13}
\BIBentryALTinterwordspacing
N.~Finn, P.~Thubert, B.~Varga, and J.~Farkas, ``{Deterministic Networking
  Architecture},'' Internet Engineering Task Force, Internet-Draft
  draft-ietf-detnet-architecture-13, May 2019, work in Progress. [Online].
  Available:
  \url{https://datatracker.ietf.org/doc/html/draft-ietf-detnet-architecture-13}
\BIBentrySTDinterwordspacing

\bibitem{rfc2210}
\BIBentryALTinterwordspacing
J.~T. Wroclawski, ``{The Use of RSVP with IETF Integrated Services},'' RFC
  2210, Sep. 1997. [Online]. Available:
  \url{https://rfc-editor.org/rfc/rfc2210.txt}
\BIBentrySTDinterwordspacing

\bibitem{finn-detnet-bounded-latency-03}
\BIBentryALTinterwordspacing
N.~Finn, J.-Y.~L. Boudec, E.~Mohammadpour, J.~Zhang, B.~Varga, and J.~Farkas,
  ``{DetNet Bounded Latency},'' Internet Engineering Task Force, Internet-Draft
  draft-finn-detnet-bounded-latency-03, Mar. 2019, work in Progress. [Online].
  Available:
  \url{https://datatracker.ietf.org/doc/html/draft-finn-detnet-bounded-latency-03}
\BIBentrySTDinterwordspacing

\bibitem{rfc7657}
\BIBentryALTinterwordspacing
D.~L. Black and P.~Jones, ``{Differentiated Services (Diffserv) and Real-Time
  Communication},'' RFC 7657, Nov. 2015. [Online]. Available:
  \url{https://rfc-editor.org/rfc/rfc7657.txt}
\BIBentrySTDinterwordspacing

\bibitem{mills1991internet}
D.~L. Mills, ``Internet time synchronization: The network time protocol,''
  \emph{IEEE Transactions on Communications}, vol.~39, no.~10, pp. 1482--1493,
  1991.

\bibitem{Huston2018time}
\BIBentryALTinterwordspacing
Is the internet running late? [Online]. Available:
  \url{https://blog.apnic.net/2018/11/28/is-the-internet-running-late/}
\BIBentrySTDinterwordspacing

\bibitem{Thangamuthu2015}
S.~Thangamuthu, N.~Concer, P.~J.~L. Cuijpers, and J.~J. Lukkien, ``Analysis of
  ethernet-switch traffic shapers for in-vehicle networking applications,'' in
  \emph{Proc. IEEE Design, Automation Test in Europe Conf. Exhibition}, Mar.
  2015, pp. 55--60.

\bibitem{thiele2015timeaware}
D.~Thiele, R.~Ernst, and J.~Diemer, ``Formal worst-case timing analysis of
  ethernet {TSN}'s time-aware and peristaltic shapers,'' in \emph{Proc. IEEE
  Vehicular Networking Conference (VNC)}, 2015, pp. 251--258.

\bibitem{ami2018sdn}
R.~Amin, M.~Reisslein, and N.~Shah, ``Hybrid {SDN} networks: A survey of
  existing approaches,'' \emph{IEEE Communications Surveys \& Tutorials},
  vol.~20, no.~4, pp. 3259--3306, 2018.

\bibitem{biz2016sdn}
N.~Bizanis and F.~A. Kuipers, ``{SDN} and virtualization solutions for the
  {Internet of Things}: A survey,'' \emph{IEEE Access}, vol.~4, pp. 5591--5606,
  2016.

\bibitem{kel2019ada}
W.~{Kellerer}, P.~{Kalmbach}, A.~{Blenk}, A.~{Basta}, M.~{Reisslein}, and
  S.~{Schmid}, ``Adaptable and data-driven softwarized networks: Review,
  opportunities, and challenges,'' \emph{Proceedings of the IEEE}, vol. 107,
  no.~4, pp. 711--731, April 2019.

\bibitem{chen-detnet-sr-based-bounded-latency-01}
\BIBentryALTinterwordspacing
M.~Chen, X.~Geng, and Z.~Li, ``{Segment Routing (SR) Based Bounded Latency},''
  Internet Engineering Task Force, Internet-Draft
  draft-chen-detnet-sr-based-bounded-latency-01, May 2019, work in Progress.
  [Online]. Available:
  \url{https://datatracker.ietf.org/doc/html/draft-chen-detnet-sr-based-bounded-latency-01}
\BIBentrySTDinterwordspacing

\bibitem{rfc8402}
\BIBentryALTinterwordspacing
C.~Filsfils, S.~Previdi, L.~Ginsberg, B.~Decraene, S.~Litkowski, and R.~Shakir,
  ``{Segment Routing Architecture},'' RFC 8402, Jul. 2018. [Online]. Available:
  \url{https://rfc-editor.org/rfc/rfc8402.txt}
\BIBentrySTDinterwordspacing

\bibitem{seaman2019pat}
M.~Seaman, ``Paternoster policing and scheduling, {Revision 2.1},'' May 2019,
  available from
  http://www.ieee802.org/1/files/public/docs2019/cr-seaman-paternoster-policing-scheduling-0519-v04.pdf,
  Last accessed May 25, 2019.

\bibitem{ball2019evaluation}
F.~Ball, K.~Basu, A.~Maqousi, and T.~Balikhina, ``Evaluation of communication
  latency for future time-critical smart grid measurement and control
  systems,'' in \emph{Proc. IEEE Int. Renewable and Sustainable Energy Conf.},
  2019, pp. 1--6.

\bibitem{mangin2019multiplexing}
C.~Mangin, ``Multiplexing method for scheduled frames in an ethernet switch,''
  Jan.~3 2019, uS Patent App. 16/064,361.

\bibitem{li2019enhanced}
Z.~Li, H.~Wan, Z.~Pang, Q.~Chen, Y.~Deng, X.~Zhao, Y.~Gao, X.~Song, and M.~Gu,
  ``An enhanced reconfiguration for deterministic transmission in
  time-triggered networks,'' \emph{IEEE/ACM Transactions on Networking}, 2019.

\bibitem{buratti2018joint}
C.~Buratti and R.~Verdone, ``Joint scheduling and routing with power control
  for centralized wireless sensor networks,'' \emph{Wireless Networks},
  vol.~24, no.~5, pp. 1699--1714, 2018.

\bibitem{said2019sdn}
S.~B.~H. Said, Q.~H. Truong, and M.~Boc, ``{SDN}-based configuration solution
  for {IEEE 802.1} time sensitive networking {(TSN)},'' \emph{ACM SIGBED
  Review}, vol.~16, no.~1, pp. 27--32, 2019.

\bibitem{wetterwald2018source}
P.~Wetterwald, P.~Thubert, and E.~M. Levy-Abegnoli, ``Source routed
  deterministic packet in a deterministic data network,'' May~31 2018, {US
  Patent App. 15/361,563}.

\bibitem{joung2019regulating}
J.~Joung, ``Regulating scheduler {(RSC)}: A novel solution for {IEEE 802.1}
  time sensitive network {(TSN)},'' \emph{Electronics}, vol.~8, no.~2, p. 189,
  2019.

\bibitem{ayub2018priority}
Q.~Ayub, A.~Ngadi, S.~Rashid, and H.~A. Habib, ``Priority queue based reactive
  buffer management policy for delay tolerant network under city based
  environments,'' \emph{PloS one}, vol.~13, no.~2, p. e0191580, 2018.

\bibitem{ma2018evaluation}
T.~Ma, W.~Hu, Y.~Wang, D.~El-Banna, J.~Grant, and H.~Dai, ``Evaluation of
  flexilink as deterministic unified real-time protocol for industrial
  networks,'' in \emph{Proc. IEEE Int. Conf. On Trust, Security And Privacy In
  Computing And Communications/IEEE Int. Conf. On Big Data Science And
  Engineering (TrustCom/BigDataSE)}, 2018, pp. 22--27.

\bibitem{mahdian2018mindelay}
M.~Mahdian and E.~Yeh, ``{MinDelay}: Low-latency joint caching and forwarding
  for multi-hop networks,'' in \emph{Proc. IEEE Int. Conf. on Commun. (ICC)},
  2018, pp. 1--7.

\bibitem{merlin2018latency}
C.~J. Merlin, L.~P. Ma, G.~Lauer, and S.~Zabele, ``Latency-aware forwarding for
  {IRON}: Latency support for back-pressure forwarding,'' in \emph{Proc. IEEE
  Military Commun. Conf. (MILCOM)}, 2018, pp. 1--6.

\bibitem{suksomboon2018configuring}
K.~Suksomboon, N.~Matsumoto, S.~Okamoto, M.~Hayashi, and Y.~Ji, ``Configuring a
  software router by the {Erlang}-$k$-based packet latency prediction,''
  \emph{IEEE Journal on Selected Areas in Communications}, vol.~36, no.~3, pp.
  422--437, 2018.

\bibitem{ngo2018efficient}
D.-M. Ngo, C.~Pham-Quoc, and T.~Ngoc~Thinh, ``An efficient high-throughput and
  low-latency {SYN} flood defender for high-speed networks,'' \emph{Security
  and Communication Networks}, vol. 2018, 2018.

\bibitem{kim2018latency}
C.~Kim, Y.~Oh, and J.~Lee, ``Latency-based graph selection manager for
  end-to-end network service on heterogeneous infrastructures,'' in \emph{Proc.
  Int. Conf. on Info. Net. (ICOIN)}.\hskip 1em plus 0.5em minus 0.4em\relax
  IEEE, 2018, pp. 534--539.

\bibitem{koutsiamanis2018best}
R.-A. Koutsiamanis, G.~Z. Papadopoulos, X.~Fafoutis, J.~M. Del~Fiore,
  P.~Thubert, and N.~Montavont, ``From best effort to deterministic packet
  delivery for wireless industrial {IoT} networks,'' \emph{IEEE Transactions on
  Industrial Informatics}, vol.~14, no.~10, pp. 4468--4480, 2018.

\bibitem{levy2019establishing}
E.~M. Levy-Abegnoli, P.~Thubert, and P.~Wetterwald, ``Establishing
  deterministic multicast paths in a network,'' Feb.~26 2019, {US Patent App.
  10/218,602}.

\bibitem{khan2018traffic}
S.~Khan, M.~Alam, M.~Fr{\"a}nzle, N.~M{\"u}llner, and Y.~Chen, ``A traffic
  aware segment-based routing protocol for {VANETs} in urban scenarios,''
  \emph{Computers \& Electrical Engineering}, vol.~68, pp. 447--462, 2018.

\bibitem{thubert2019deterministic}
P.~Thubert, P.~Wetterwald, R.~Ramachandran, and E.~M. Levy-Abegnoli,
  ``Deterministic stitching of deterministic segments across distinct
  deterministic domains,'' Mar.~28 2019, {US Patent App. 15/713,827}.

\bibitem{kim2018kreonet}
D.~Kim, Y.-H. Kim, C.~Park, and K.-I. Kim, ``{KREONET-S}: Software-defined wide
  area network design and deployment on {KREONET},'' \emph{IAENG International
  Journal of Computer Science}, vol.~45, no.~1, 2018.

\bibitem{grant2018new}
J.~Grant, ``New packet routing for {5G} to replace {TCP/IP},'' in \emph{Audio
  Engineering Society Convention 144}.\hskip 1em plus 0.5em minus 0.4em\relax
  Audio Engineering Society, 2018.

\bibitem{chen2019etn}
Q.~Chen, Q.~Zhong, D.~Sun, R.~Li, L.~Niu, and L.~Ding, ``{ETN}--ethernet
  transport network for {5G} mobile transport, metro, and {DCI} network,'' in
  \emph{Proc. IEEE Int. Conf. on Communication Sys.}, 2019, pp. 332--336.

\bibitem{pointurier2019end}
Y.~Pointurier, N.~Benzaoui, W.~Lautenschlaeger, and L.~Dembeck, ``End-to-end
  time sensitive optical networking: Challenges and solutions,'' \emph{IEEE/OSA
  Journal of Lightwave Technology}, vol.~37, no.~7, pp. 1732--1741, Apr. 2019.

\bibitem{bocquillon2018robust}
R.~Bocquillon and A.~Jouglet, ``Robust routing in deterministic delay-tolerant
  networks,'' \emph{Computers \& Operations Research}, vol.~92, pp. 77--86,
  2018.

\bibitem{borah2018energy}
S.~J. Borah, S.~K. Dhurandher, I.~Woungang, N.~Kandhoul, and J.~J. Rodrigues,
  ``An energy-efficient location prediction-based forwarding scheme for
  opportunistic networks,'' in \emph{Proc. IEEE Int. Conf. on Commun. (ICC)},
  2018, pp. 1--6.

\bibitem{jalan2018forwarding}
R.~Jalan, G.~Kamat, S.~Sankar, and H.~Karampurwala, ``Forwarding data packets
  using a service-based forwarding policy,'' Apr.~10 2018, {US Patent
  9,942,152}.

\bibitem{thubert2019transport}
P.~Thubert, P.~Wetterwald, and E.~M. Levy-Abegnoli, ``Transport layer providing
  deterministic transport across multiple deterministic data links,'' Apr.~11
  2019, {US Patent App. 15/725,502}.

\bibitem{thubert2018bit}
P.~Thubert, J.-P. Vasseur, P.~Wetterwald, and I.~Wijnands, ``Bit indexed
  explicit replication for deterministic network data plane,'' Nov.~27 2018,
  {US Patent App. 10/142,227}.

\bibitem{wetterwald2018insertion}
P.~Wetterwald, P.~Thubert, E.~M. Levy-Abegnoli, and J.-P. Vasseur, ``Insertion
  of management packet into a deterministic track,'' Jun.~5 2018, {US Patent
  9,992,703}.

\bibitem{bashandy2019segment}
A.~R. Bashandy, C.~Filsfils, and D.~D. Ward, ``Segment routing over label
  distribution protocol,'' Apr.~23 2019, {US Patent App. 10/270,664}.

\bibitem{filsfils2019scalable}
C.~Filsfils, R.~Gandhi, T.~Saad, S.~Soni, and P.~Khordoc, ``Scalable
  distributed end-to-end performance delay measurement for segment routing
  policies,'' Mar.~12 2019, {US Patent App. 10/230,605}.

\bibitem{laberge2018tactical}
T.~LaBere, C.~Filsfils, and P.~J.~R. Francois, ``Tactical traffic engineering
  based on segment routing policies,'' May~10 2018, {US Patent App.
  15/345,049}.

\bibitem{katsalis2018towards}
K.~Katsalis, L.~Gatzikis, and K.~Samdanis, ``Towards slicing for transport
  networks: The case of {Flex-Ethernet} in {5G},'' in \emph{Prof. IEEE Conf. on
  Standards for Communications and Networking}, 2018, pp. 1--7.

\bibitem{chu2018pre}
U.~{Chunduri}, A.~{Clemm}, and R.~{Li}, ``{Preferred Path Routing} - a
  next-generation routing framework beyond {Segment Routing},'' in \emph{Proc.
  IEEE Global Commun. Conf. (GLOBECOM)}, Dec 2018, pp. 1--7.

\bibitem{eckert2018preferred}
T.~Eckert, Y.~Qu, and U.~Chunduri, ``Preferred path routing {(PPR)}
  graphs-beyond signaling of paths to networks,'' in \emph{Proc. IEEE Int.
  Conf. on Network and Service Management}, 2018, pp. 384--390.

\bibitem{wang2019sdn}
Y.-C. Wang, Y.-D. Lin, and G.-Y. Chang, ``{SDN}-based dynamic multipath
  forwarding for inter--data center networking,'' \emph{Int. Journal of
  Communication Systems}, vol.~32, no.~1, pp. e3843--1--e3843--15, Jan. 2019.

\bibitem{zhouinsight}
Z.~Zhou, M.~S. Berger, S.~R. Ruepp, and Y.~Yan, ``Insight into the {IEEE
  802.1Qcr} asynchronous traffic shaping in time sensitive network,''
  \emph{Advances in Science, Technology and Engineering Systems Journal},
  vol.~4, no.~1, pp. 292--301, 2019.

\bibitem{zho2018ana}
Z.~Zhou, Y.~Yan, M.~Berger, and S.~Ruepp, ``Analysis and modeling of
  asynchronous traffic shaping in time sensitive networks,'' in \emph{Proc.
  IEEE Int. Workshop on Factory Commun. Systems (WFCS)}, Jun. 2018, pp. 1--4.

\bibitem{dou2017ins}
A.~Douik, S.~Sorour, T.~Y. Al-Naffouri, and M.~S. Alouini, ``Instantly
  decodable network coding: From centralized to device-to-device
  communications,'' \emph{IEEE Communications Surveys \& Tutorials}, vol.~19,
  no.~2, pp. 1201--1224, Second Qu. 2017.

\bibitem{gab2018cat}
F.~Gabriel, S.~Wunderlich, S.~Pandi, F.~H. Fitzek, and M.~Reisslein,
  ``Caterpillar {RLNC} with feedback {(CRLNC-FB)}: Reducing delay in selective
  repeat {ARQ} through coding,'' \emph{IEEE Access}, vol.~6, pp.
  44\,787--44\,802, 2018.

\bibitem{tou2011fly}
P.~U. Tournoux, E.~Lochin, J.~Lacan, A.~Bouabdallah, and V.~Roca, ``On-the-fly
  erasure coding for real-time video applications,'' \emph{IEEE Transactions on
  Multimedia}, vol.~13, no.~4, pp. 797--812, 2011.

\bibitem{wun2017cat}
S.~Wunderlich, F.~Gabriel, S.~Pandi, F.~H. Fitzek, and M.~Reisslein,
  ``Caterpillar {RLNC (CRLNC)}: A practical finite sliding window {RLNC}
  approach,'' \emph{IEEE Access}, vol.~5, pp. 20\,183--20\,197, 2017.

\bibitem{cao2019dyn}
H.~Cao, H.~Zhu, and L.~Yang, ``Dynamic embedding and scheduling of service
  function chains for future {SDN/NFV}-enabled networks,'' \emph{IEEE Access},
  vol.~7, pp. 39\,721--39\,730, 2019.

\bibitem{sha2018lay}
P.~Shantharama, A.~S. Thyagaturu, N.~Karakoc, L.~Ferrari, M.~Reisslein, and
  A.~Scaglione, ``{LayBack: SDN} management of multi-access edge computing
  {(MEC)} for network access services and radio resource sharing,'' \emph{IEEE
  Access}, vol.~6, pp. 57\,545--57\,561, 2018.

\bibitem{xia2019red}
Z.~Xiang, F.~Gabriel, E.~Urbano, G.~T. Nguyen, M.~Reisslein, and F.~H. Fitzek,
  ``Reducing latency in virtual machines: Enabling tactile internet for
  human-machine co-working,'' \emph{IEEE Journal on Selected Areas in
  Communications}, vol.~37, no.~5, pp. 1098--1116, 2019.

\bibitem{yur2018ope}
M.~Yurchenko, P.~Cody, A.~Coplan, R.~Kennedy, T.~Wood, and K.~Ramakrishnan,
  ``{OpenNetVM}: A platform for high performance {NFV} service chains,'' in
  \emph{Proc. ACM of the Symposium on SDN Research}, 2018, p.~21.

\bibitem{bar2014nov}
M.~Barcelo, A.~Correa, X.~Vilajosana, J.~L. Vicario, and A.~Morell, ``Novel
  routing approach for the {TSCH} mode of {IEEE 802.15.14e} in wireless sensor
  networks with mobile nodes,'' in \emph{Proc. IEEE Vehicular Technology
  Conference (VTC2014-Fall)}, 2014, pp. 1--5.

\bibitem{guc2017uni}
J.~W. Guck, A.~Van~Bemten, M.~Reisslein, and W.~Kellerer, ``Unicast {QoS}
  routing algorithms for {SDN}: A comprehensive survey and performance
  evaluation,'' \emph{IEEE Communications Surveys \& Tutorials}, vol.~20,
  no.~1, pp. 388--415, First Qu. 2018.

\bibitem{thubert2019insertion}
P.~Thubert, P.~Wetterwald, J.-P. Vasseur, and E.~M. Levy-Abegnoli, ``Insertion
  slots along deterministic track for movable network device in a deterministic
  network,'' Jan.~10 2019, {US Patent App. 16/107,759}.

\bibitem{filsfils2019network}
C.~Filsfils and R.~Gandhi, ``Network programming for performance and liveness
  monitoring in segment routing networks,'' 2019, technical Disclosure Commons,
  https://www.tdcommons.org/dpubs\_series/2165, Last accessed 05/19/2019.

\bibitem{jadin2019cg4sr}
M.~Jadin, F.~Aubry, P.~Schaus, and O.~Bonaventure, ``{CG4SR}: Near optimal
  traffic engineering for segment routing with column generation,'' in
  \emph{IEEE Infocom. Proc.}, 2019.

\end{thebibliography}


\end{document}